\documentclass[letterpaper, 10 pt, conference]{ieeeconf}

\IEEEoverridecommandlockouts                              

\usepackage{grffile}
\usepackage{caption}
\usepackage{graphicx}
\usepackage{amssymb,amsmath}
\usepackage{epsfig}
\usepackage{epstopdf}
\usepackage{setspace}
\usepackage{amsthm}
\usepackage{color}
\usepackage{algorithm}
\usepackage{algcompatible}
\usepackage{float}
\usepackage[english]{babel}
\usepackage[autostyle]{csquotes}
\usepackage{algpseudocode}
\usepackage{enumerate}
\usepackage{tikz}
\usepackage{multirow}
\usetikzlibrary{shapes,arrows}
\title{\LARGE \bf
A Distributed Nash Equilibrium Seeking in Networked Graphical Games
}

\author{Farzad Salehisadaghiani, and Lacra Pavel
\thanks{The authors are with the Department of Electrical and Computer Engineering, University of Toronto, Toronto, ON M5S 3G4, Canada (e-mails: {\tt\small farzad.salehisadaghiani@mail.utoronto.ca, pavel@ece.utoronto.ca}).}}
\newtheorem{assumption}{Assumption}
\newtheorem{lemma}{Lemma}
\newtheorem{theorem}{Theorem}
\newtheorem{remark}{Remark}

\newtheorem{definition}{Definition}

\newtheorem{corollary}{Corollary}

\begin{document}
\allowdisplaybreaks

\maketitle
\thispagestyle{empty}
\pagestyle{empty}

\color{black}\begin{abstract}
This paper considers  a distributed gossip approach for finding a Nash equilibrium in networked games on graphs.
In such games a player's cost function may be affected by the actions of any subset of players. An interference graph is employed to illustrate the partially-coupled cost functions and the asymmetric information requirements. For a given interference graph,  network communication between players  is considered to be limited.  A generalized communication graph  is designed so that players exchange only their required information.  An algorithm is designed whereby  players, with possibly partially-coupled cost functions, make decisions based on the estimates of other players' actions obtained from local neighbors.  It is  shown that this choice of communication graph guarantees that all players' information is exchanged after sufficiently many iterations. Using a set of standard assumptions on the cost functions, the  interference and the communication graphs,  almost sure convergence to a Nash equilibrium is proved for diminishing step sizes. Moreover, the case when the cost functions are not known by the players is investigated and a convergence proof is presented for diminishing step sizes. The effect of the second largest eigenvalue of the expected communication matrix on the convergence rate is quantified. The trade-off between parameters associated with the communication graph and the ones associated with the interference graph is illustrated. Numerical results are presented for a large-scale networked game.
\end{abstract}
\color{black}\section{INTRODUCTION}

Distributed seeking of Nash equilibria in networked games has received considerable attention in recent years \cite{stankovic2012distributed}-\cite{Marden2013}. A networked game can be represented by a \emph{graphical model} where  the cost function of each player can be indexed as a function of player's own actions and those of his neighbors in the graph. There are many real-world applications that motivate us to generalize the Nash seeking problem to a graphical game setup \cite{li2010competitive}, \cite{chen2012spatial}. For instance, the collection of transmitters and receivers in a wireless data network can be described by a graphical model. Interferences among the transmitters and receivers affect the players' signal-to-interference ratio (SIR) \cite{alpcan2004hybrid}. Another relevant application that can be modeled as a graphical game is optical network. The channels are assumed to be the players and interferences, which affect the optical signal-to-noise ratio (OSNR) of each channel, can be modeled by graph edges, \cite{Pavelbook2012}.

In this work  a locally distributed algorithm is designed towards  Nash equilibrium seeking in a graphical game. In such a game, the players' cost functions may {\color{black}depend on} the actions of \emph{any subset} of players. Players exchange the required information locally according to a communication graph and update their actions to optimize their cost functions. With the limited information available from local neighbors, each player maintains an estimate of the other players' actions and update their estimates over time.

\emph{Literature review.} A graphical game is a succinct representation of a multi-player game which considers the local interactions and the sparsity of the interferences.  Such a  game can be simply described by an undirected graph called \emph{interference graph} in which the players are marked by the vertices and the interferences are represented by the edges \cite{nisan2007algorithmic}, \cite{kearns2001graphical}.

The idea of a graphical game has been used in various areas. In congestion games, \cite{tekin2012atomic} considers a generalization to graphical games. The model involves the spatial positioning of the players which affects their performances. A \emph{conflict graph} is defined to specify the players that cause congestion to each other. In \cite{Marden2013} a methodology is presented  for games with local cost functions which are dependent on information from only a set of local neighboring agents. Extra state space variables are defined for the game to achieve a desired degree of locality. In \cite{abouheaf2014multi}, graphical games are considered in the context of dynamical games, where the dynamic of each player depends only on local neighbor information. A stronger definition of an interactive Nash equilibrium is used to guarantee a unique Nash equilibrium. Moreover, the information flow is described by a communication graph which is \emph{identical} to the interference graph. In an economic setting, \cite{bramoulle2014strategic} draws attention to the problem of \enquote{who interacts with whom} in a network. This paper states the importance of communication with neighboring players in the network. The effect of local peers on increasing the usage level of consumers is addressed in \cite{candogan2012optimal}. Using word-of-mouthcommunication, players typically form their opinions about the quality of a product and improve their purchasing behavior based on the information otained from local peers.

In \cite{zhu2016distributed} the problem of finding a Nash equilibrium is studied for generalized convex games. The interference graph may not necessarily  be a complete graph, but the communication graph is \emph{identical} to the interference graph. A connected communication graph is considered in \cite{Jayash} for the class of aggregative games. In this game, it is assumed that the interferences on each player originate from all other players in the network which is literally the case with a \emph{complete interference graph}. For a large class of convex games, \cite{Salehisadaghiani2014Nash} proposes an asynchronous gossip-based algorithm over a connected communication graph. {\color{black}A complete interference graph specifies the interferences between the players. The algorithm is based on projected gradient method that uses diminishing step sizes.} {\color{black}Thereafter, the algorithm is extended in \cite{salehisadaghiani2016distributed} for the case with constant step sizes and it is proved that the algorithm can locate a small neighborhood of a Nash equilibrium of the game over a complete interference graph.

This work is also related to the literature on distributed optimization \cite{nedic2011asynchronous,johansson2008distributed,nedic2009distributed}.
In a distributed optimization problem, agents, who communicate over a connected graph, minimize an aggregate of the cost functions with respect to a common optimization variable. The method that is used in \cite{bertsekas2003convex,nedic2011asynchronous} is to incrementally update the optimization variable by each agent using the gradient information corresponding to a single component function of that agent. Then, the updated variable is passed to other agents and this process is repeated until they reach a consensus which is an optimal point of this problem. The convergence rate of these algorithms is shown to be tightly dependent on the spectral gap of the underlying communication graph \cite{duchi2012dual}. While the techniques we use here are similar, there are technical difficulties due to the game context. Unlike the distributed optimization case where each agent updates/controls the local copy of the decision vector, in a game context each player controls his action which is only an element of the decision vector. However we circumvent this problem by assuming an estimate of the other players' decisions and update them by the received information from the communication.

\emph{Contributions.} In this work, we propose a gossip-based algoruthm to find a Nash equilibrium of networked games.} We generalize the algorithm in \cite{Salehisadaghiani2014Nash,salehisadaghiani2016distributed} to the case when the interference graph is not a complete graph, i.e., when the players' cost functions are affected by the actions of any subset of players. {\color{black}Thus, each player maintains only an estimate of the players' actions that interfere with his cost function.} Communications are assumed to be limited and a communication graph is considered to be a subset of the interference graph. {\color{black}Since, each player may maintain an estimate of different players' actions based on the interference graph, the communication graph needs to be designed in a way that all the players obtain the required information from the neighbors to update their estimates.} We prove that there exists a lower bound for the communication graph under which the algorithm converges to a Nash equilibrium for diminishing step sizes. We then discuss the case when the cost functions (models) are not available to the players but only the realized cost values at the certain points are. Using finite-difference technique to approximate the gradient, we present an almost-sure convergence to a Nash equilibrium. This method has been used in \cite{zhu2016distributed} to approximate the gradient of cost functions leading toward a gradient-free algorithm to compute a Nash equilibrium. This is referred to as adaptiveness property in \cite{zhu2016distributed} when the algorithm is able to compute equilibrium point despite the lack of game components (e.g., cost functions) due to system policy or the national security.

Lastly, inspired by \cite{boyd2006randomized}, we investigate the convergence rate of the proposed algorithm. \cite{boyd2006randomized} show that the convergence time of an averaging algorithm under gossip constraint is dependent on the second largest eigenvalue of a doubly stochastic matrix characterizing the algorithm (See also \cite{duchi2012dual,olshevsky2009convergence}). We prove that the convergence rate diminishes as the second largest eigenvalue of the expected communication matrix grows. The results show a relation between this parameter and the ones associated with the communication and the interference graphs. The trade-off between the parameters associated with these two graphs is illustrated.

The paper is organized as follows. The problem statement and assumptions are provided in Section~II. A locally distributed algorithm is proposed in Section~III. The convergence of the algorithm with diminishing step sizes is discussed in Section~IV, while the non-model based approach is investigated in Section~V. The convergence rate analysis is then presented in Section~VI. Simulation results are demonstrated in Section~VII.
\subsection{Notations and Notions} 

All vector norms $\|\cdot\|$ are Euclidean. The cardinality of a set $A$ is denoted by $|A|$. The Euclidean projection of $x$ onto the set $K$ is denoted by $T_K[x]$. We denote by $[a_{ij}]_{i,j=1,\ldots,N}$ an $N\times N$ matrix with $a_{ij}$ as the entry of the $i$-th row and the $j$-th column. We also denote by $[a_i]_{i=1,\ldots,N}$ an $N\times 1$ vector with $a_i$ as the $i$-th entry. The $N\times N$ identity matrix is denoted by $I_N$. We denote by $\textbf{1}_N$ an $N\times 1$ vector whose entries are all equal to 1 and by $\textbf{0}_N$ an $N\times 1$ vector whose entries are all equal to 0. We use $e_i$ to denote a unit vector whose $i$-th element is 1 and the others are 0.

The following definitions are from \cite{godsil2013algebraic}, \cite{goddard1993note}. An undirected graph $G$, or simply graph, is a pair $(V,E)$ with $V$ as a finite set of vertices and $E\subseteq V\times V$ a set of edges such that for $i,j\in V$, if $(i,j)\in E$, then $(j,i)\in E$. The degree of vertex $i$, denoted by $\text{deg}(i)$, is the number of edges connected to $i$. A path in a graph is a sequence of edges which connects a sequence of vertices. A graph is connected if there is a path between every pair of vertices. An adjacency matrix $A=[a_{ij}]_{i,j\in V}$ is a matrix with $a_{ij}=1$ if $(i,j)\in E$ and $a_{ij}=0$ otherwise.

A subgraph $H$ of a graph $G$ is a graph whose vertices and edges are a subset of the vertex and edge set of $G$, respectively. A supergraph $H$ of  $G$ is a graph of which $G$ is a subgraph. A subgraph $H$ is a spanning subgraph of $G$, if it contains all the vertices of $G$. A triangle-free spanning subgraph $H$ of  $G$ is a subgraph in which no three vertices form a triangle of edges. {\color{black}Moreover, $H$ is a maximal triangle-free spanning subgraph of $G$ if adding an edge from $G-H$ to $H$ creates only one triangle. Note that if $G$ has no triangle, the maximal triangle-free spanning subgraph $H$ becomes identical to $G$.}
\color{black}\section{Problem Statement}\label{problem_statement}

Consider a multi-player game in a network with a set of players $V=\{1,\ldots,N\}$. Each player $i\in V$ has  a real-valued cost function $J_i$. Players' cost functions are not necessarily fully coupled in the sense that they may be affected by the actions of any number of players. To illustrate the partially coupled cost functions, we define an \emph{interference graph}, denoted by $G_I(V,E_I)$\footnote[1]{In this paper, we assume that $G_I$ is not a complete graph.}, with $E_I$ marking player pairs that interfere one with another. We denote with $N_I(i)$, the set of neighbors of player $i$ in $G_I$, i.e., $N_I(i):=\{j\in V|(i,j)\in E_I\}$. We also define $\tilde{N}_I(i):=N_I(i)\cup\{i\}$.
\begin{assumption}\label{connected_undirected}
The interference graph $G_I$ is connected and undirected.
\end{assumption}
Let $\Omega_j\subset\mathbb{R}$ denote the action set of player $j$. We denote by $\Omega$ the action set of all players, i.e., $\Omega=\prod_{i\in V}\Omega_i\subset\mathbb{R}^N$ where $\prod$ denotes the Cartesian product. For $i\in V$, $J_i:\Omega^i\rightarrow \mathbb{R}$ is the cost function of player $i$ where $\Omega^i=\prod_{j\in\tilde{N}_I(i)}\Omega_j\subset\mathbb{R}^{|\tilde{N}_I(i)|}$ is the action set of players interfering with the cost function of player $i$. The game denoted by $\mathcal{G}(V,\Omega_i,J_i, G_I)$ is defined based on the set of players $V$, the action set $\Omega_i$ $\forall i\in V$, the cost function $J_i$ $\forall i\in V$ and $G_I$.
For $i\in V$, let $x^i=(x_i,x_{-i}^i)\in\Omega^i$, with $x_i\in\Omega_i$ and $x_{-i}^i\in\Omega_{-i}^i:=\prod_{j\in N_I(i)}\Omega_j$, denote the other players' actions which interfere with the cost function of player $i$. Let also $x=(x_i,x_{-i})\in\Omega$, with $x_i\in\Omega_i$ and $x_{-i}\in\Omega_{-i}:=\prod_{j\in V/\{i\}}\Omega_j$, denote all other players' actions except $i$.

The game defined on $G_I$ is played such that for given $x_{-i}^i\in \Omega_{-i}^i$, each player $i$ aims to minimize his own cost function selfishly to find an optimal action,
\begin{equation}
\label{mini_0}
\begin{aligned}
& \underset{y_i}{\text{minimize}}
& & J_i(y_i,x_{-i}^i) \\
& \text{subject to}
& & y_i\in \Omega_i.
\end{aligned}
\end{equation}
Note that there are $N$ separate simultaneous optimization problems and each of them is run by a particular player $i$. We assume that the cost function $J_i$ and the action set $\Omega^i$ are only available to player $i$.  Thus every player knows which other players' actions affect his cost function.

A Nash equilibrium for the case when $G_I$ is not a complete graph is defined as follows.
\begin{definition}\label{Nash_def}
Consider an $N$-player game $\mathcal{G}(V,\Omega_i,J_i, G_I)$, each player $i$ minimizing the cost function $J_i:\Omega^i\rightarrow\mathbb{R}$. A vector $x^*=(x_i^*,x_{-i}^*)\in\Omega$ is called a Nash equilibrium of this game if for every given $x_{-i}^{i*}\in\Omega_{-i}^i$
\begin{equation}
J_i(x_i^*,{x_{-i}^{i*}})\leq J_i(x_{i},{x_{-i}^{i*}})\quad\forall x_i\in \Omega_i,\,\,\forall i\in V.
\end{equation}
\end{definition}
Definition~\ref{Nash_def} is a restatement of a Nash equilibrium definition so that  when $G_I$ is not a complete graph, $J_i(x_i,x_{-i})$ and $J_i(x_i^*,x_{-i}^*)$ are replaced with $J_i(x_i,x_{-i}^i)$ and $J_i(x_i^*,{x_{-i}^{i*}})$, respectively.

We assume that players exchange some information in order to update their actions. A \emph{communication graph} $G_C(V,E_C)$ is defined where $E_C\subseteq V\times V$ denotes the set of communication links between the players. $(i,j)\in E_C$ if and only if players $i$ and $j$ communicate together. The set of neighbors of player $i$ in $G_C$, denoted by $N_C(i)$, is defined as $N_C(i):=\{j\in V|(i,j)\in E_C\}$. In order to reduce the number of communications between the players, we design an assumption on $G_C$ in such a way that only the required information is obtained by the players. {\color{black}Particularly, for each player $i$, the required information that needs to be obtained is $\{x_j:j\in N_I(i)\}$.}

{\color{black}
	Let $G_m$ be a maximal triangle-free spanning subgraph of $G_I$. Then we have the following assumption for $G_C$.
	
	\begin{assumption}\label{connectivity}
	The communication graph $G_C$ satisfies
	\begin{equation*}
		G_m\subseteq G_C\subseteq G_I.
	\end{equation*}
\end{assumption}
\begin{remark}
	Note that the maximal triangle-free subgraph $G_m$ is only a lower bound for $G_C$ (in other words, $G_m$ is a sparsest possible $G_C$). If $G_m$ does not exist, Assumption~\ref{connectivity} could be replaced by the following condition:
	\begin{itemize}
		\item Check that for every $i\in V$ and $j\in N_I(i)$, there is a path of length 1 or 2 between $i$ and $j$ in $G_C$.
	\end{itemize}
\end{remark}}
\begin{remark}\label{G_m_not_unique}
$G_m$ is not unique in the sense that any selection of a maximal triangle-free subgraph of $G_I$ could be considered in Assumption~\ref{connectivity}. Since $G_m$ is connected and undirected, by Assumption~\ref{connectivity}, $G_C$ is connected and undirected.
\end{remark}
A Nash equilibrium can be characterized in terms of a variational inequality problem for  the pseudo-gradient mapping  $F:\Omega\rightarrow\mathbb{R}^N$,
\begin{equation}
F(x):=[\nabla_{x_i}J_i(x^i)]_{i\in V},
\end{equation}
as in the following lemma (Proposition~1.5.8, page~83 in  \cite{Facchi24}).
\begin{lemma}
\label{Nash_lemma}
$x^*$ is a Nash equilibrium of the game represented by \eqref{mini_0} if and only if
\begin{equation}\label{NE_char}
x^*=T_{\Omega}[x^*-\alpha {F}(x^*)]
\end{equation}
for $\alpha>0$, where  $T_\Omega:\mathbb{R}^N\rightarrow\Omega$ is an Euclidean projection. 
\end{lemma}
We state a few assumptions for the existence and the uniqueness of a Nash equilibrium.
\begin{assumption}
\label{assump}
For every $i\in V$, the action set $\Omega_i$ is a non-empty, compact and convex subset of $\mathbb{R}$. $J_i(x_i,x_{-i}^i)$ is a continuously differentiable function in $x_i$, jointly continuous in $x^i$ and convex in $x_i$ for every $x_{-i}^i$.
\end{assumption}
The compactness of $\Omega$ implies that $\forall i\in V$ and $x^i\in\Omega^i$, 
\begin{equation}\label{bounded}
\|\nabla_{x_i}J_i(x^i)\|\leq C,\quad\text{for some }C>0.
\end{equation}
\begin{assumption}\label{Lip_assump}
$F:\Omega\rightarrow\mathbb{R}^N$ is strictly monotone,
\begin{equation}
(F(x)-F(y))^T(x-y)> 0\quad\forall x,y\in \Omega,\text{ }x\neq y.
\end{equation}
\end{assumption}
{\color{black}Note that the strict monotonicity of $F$ implies the uniqueness of Nash equilibrium.}
\begin{assumption}\label{Lip_assump2}
$\nabla_{x_i}J_i(x_i,u)$ is Lipschitz continuous in $x_i$, for every fixed $u\in\Omega_{-i}^i$ and for every $i\in V$, i.e., there exists $\sigma_i>0$ such that
\begin{equation}
\|\nabla_{x_i}J_i(x_i,u)-\nabla_{x_i}J_i(y_i,u)\|\leq \sigma_i\|x_i-y_i\|\quad\forall x_i,y_i\in\Omega_{i}.
\end{equation}
Moreover, $\nabla_{x_i}J_i(x_i,u)$ is Lipschitz continuous in $u$ with a Lipschitz constant $L_i>0$ for every fixed $x_i\in\Omega_i,\,\forall i\in V$.
\end{assumption}
\begin{remark}\label{Lipscitz_F}
Assumption~\ref{Lip_assump2} implies that $\nabla_{x_i}J_i(x^i)$ is a Lipschitz continuous in $x^i\in\Omega^i$ with a Lipschitz constant $\rho_i=\sqrt{2L_i^2+2\sigma_i^2}$ for every $i\in V$. Moreover, $F(x)$ is also Lipschitz continuous in $x\in\Omega$ with a Lipschitz constant $\rho=\sqrt{\sum_{i\in V}\rho_i^2}$.
\end{remark}
Our objective is to find an algorithm for computing a Nash equilibrium of $\mathcal{G}(V,\Omega_i,J_i, G_I)$ with partially coupled cost functions as described by $G_I(V,E_I)$ using only imperfect information over the communication graph $G_C(V,E_C)$.
\color{black}\section{Asynchronous Gossip-based Algorithm}\label{asynch}
We propose a distributed algorithm, using an asynchronous gossip-based method in \cite{Salehisadaghiani2014Nash}. We obtain a Nash equilibrium of $\mathcal{G}(V,\Omega_i,J_i, G_I)$ by solving the associated $VI$ problem by a projected gradient-based approach with diminishing step size.
The mechanism of the algorithm can be briefly explained as follows:
Each player builds and maintains an estimate of the actions which interfere with his cost function specified by $G_I$ and locally communicates with his neighbors over $G_C$ to exchange his estimates and update his action. The algorithm is inspired by \cite{Salehisadaghiani2014Nash} except that only the required information is exchanged according to $G_I$. Thus, when $G_I$ is not complete, the proposed algorithm can offer substantial savings. The convergence proof depends on a generalized weight matrix, whose properties need to be investigated and proved.

The algorithm is elaborated in the following steps:\\
1- \textbf{\emph{Initialization Step:}}
Each player $i$ maintains an initial \emph{temporary} estimate $\tilde{x}^i(0)\in\Omega^i$ for the players whose actions interfere with his cost function. Let $\tilde{x}_j^i(0)\in\Omega_j\subset\mathbb{R}$ be player $i$'s initial temporary estimate of player $j$'s action, for $i\in V,\,j\in\tilde{N}_I(i)$. Then, $\tilde{x}^i(0)=[\tilde{x}^i_j(0)]_{j\in\tilde{N}_I(i)}$.\\
2- \textbf{\emph{Gossiping Step:}}
At the gossiping step, player $i_k$ wakes up at $T(k)$ {\color{black}uniformly at random} and selects a communication neighbor {\color{black}with an equal probability} indexed by $j_k\in N_C(i_k)$.
They exchange their temporary estimate vectors and construct their final estimates. Let $\tilde{x}_j^i(k)\in\Omega_j\subset\mathbb{R}$ be player $i$'s temporary estimate of player $j$'s action at $T(k)$. Then he constructs his estimate $\hat{x}^i(k)\in\Omega^i$ of the players whose actions interfere with his cost function. Let $\hat{x}_j^{i}(k)\in\Omega_j\subset\mathbb{R}$ be player $i$'s estimate of player $j$'s action, for $i\in V,\,j\in\tilde{N}_I(i)$.

The estimates are computed as in the following:
\begin{eqnarray}\label{excluding}
\hspace{0cm}\text{1)} \begin{cases}
\hat{x}_{l}^{i_k}(k)=\frac{\tilde{x}_{l}^{i_k}(k)+\tilde{x}_{l}^{j_k}(k)}{2},& l\in (N_I(i_k)\cap\tilde{N}_I(j_k)) \\ \hat{x}_{l}^{j_k}(k)=\frac{\tilde{x}_{l}^{i_k}(k)+\tilde{x}_{l}^{j_k}(k)}{2},& l\in (N_I(j_k)\cap\tilde{N}_I(i_k)).
\end{cases}
\end{eqnarray}
\begin{eqnarray}\label{including}
\hspace{0cm}\text{2)} \begin{cases}
\hat{x}_{r}^{i_k}(k)=\tilde{x}_{r}^{i_k}(k),&r\in \tilde{N}_I(i_k)\backslash (N_I(i_k)\cap\tilde{N}_I(j_k))\\
\hat{x}_{r}^{j_k}(k)=\tilde{x}_{r}^{j_k}(k),&r\in \tilde{N}_I(j_k)\backslash (N_I(j_k)\cap\tilde{N}_I(i_k)).
\end{cases}
\end{eqnarray}\\
3) For all other $i\notin \{i_k,j_k\}$,
\begin{equation}\label{other_inc_exc}
\hat{x}_j^i(k)=\tilde{x}_j^i(k),\quad\forall i\notin \{i_k,j_k\},\,\forall j\in\tilde{N}_I(i).
\end{equation}
Note that  $\tilde{x}_i^i(k)=x_i(k)$ for all $i\in V$, since no estimation is needed for its own action.

{\color{black}By the following lemma, we show that to update the temporary estimates and construct the final estimates, each player $i\in V$ obtains all necessary information about the players in $N_I(i)$.
\begin{lemma}\label{tria_lemma}
		Let $G_I$ and $G_C$ satisfying Assumptions~\ref{connected_undirected} and \ref{connectivity}. Then $\forall i\in V$,
		\begin{equation}\label{Lemma_set}
		\bigcup_{j\in N_C(i)}\big(N_I(i)\cap\tilde{N}_I(j)\big)=N_I(i).
		\end{equation}
\end{lemma}
\par{\emph{Proof}}. See Appendix.
\begin{remark}
	By Lemma~\ref{tria_lemma}, all the information obtained by player $i$ from $\forall j\in N_C(i)$, i.e., $\bigcup_{j\in N_C(i)}\big(N_I(i)\cap\tilde{N}_I(j)\big)$ is equal to the necessary information that this player needs to update his estimates.
\end{remark}}
\begin{remark}
	We show via a counter example that the bound $G_m$ is needed, since otherwise, if players communicate via a path of length greater than 2, they may lose some information. Consider a 4-player game in a network with the interference graph $G_I$ and the communication graph $G_C$ as in Fig.~1.
	\begin{figure}[!htb]\label{ax_G_I}
		\centering
		\begin{tikzpicture}
		\fill (1/2,1/2) circle (2pt);
		\fill (-1/2,1/2) circle (2pt);
		\fill (-1/2,-1/2) circle (2pt);
		\fill (1/2,-1/2) circle (2pt);
		\draw[thick] (-1/2,1/2) node[above left, scale=0.7] {$1$} --
		(1/2,1/2) node[above right, scale=0.7] {$2$} --
		(1/2,-1/2) node[below right, scale=0.7] {$3$} --
		(-1/2,-1/2) node[below left, scale=0.7] {$4$};
		\draw[thick] (-1/2,-1/2) -- (-1/2,1/2);
		\draw[thick] (-1/2,1/2) -- (1/2,-1/2);
		\end{tikzpicture}
		\hspace{2cm}
		\begin{tikzpicture}
		\fill (1/2,1/2) circle (2pt);
		\fill (-1/2,1/2) circle (2pt);
		\fill (-1/2,-1/2) circle (2pt);
		\fill (1/2,-1/2) circle (2pt);
		\draw[thick] (-1/2,1/2) node[above left, scale=0.7] {$1$} --
		(1/2,1/2) node[above right, scale=0.7] {$2$} --
		(1/2,-1/2) node[below right, scale=0.7] {$3$};
		\draw[thick] (-1/2,-1/2) node[below left, scale=0.7] {$4$} -- (-1/2,1/2);
		\end{tikzpicture}
		\caption{(a) Interference graph $G_I$ (b) Communication graph $G_C$.}
	\end{figure}
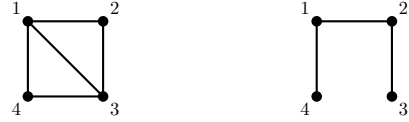
	In this example $G_C\nsupseteq G_m$. Note that players 3 and 4 do not have direct communication but through a path of length 3 via players 2 and 1. Let player 3 communicate with player 2. According to $G_I$, the cost functions of player 2 and 3 are $J_2(x_2,x_1,x_3)$ and $J_3(x_3,x_1,x_2,x_4)$, respectively. Since $x_4$ {\color{black}does not interfere with} the cost function of player 2, player 3 cannot obtain any information about player 4 from player 2.
\end{remark}
3- \textbf{\emph{Local Step}}

At this moment all the players update their actions according to a projected gradient-based method. Let $\hat{x}^i=(\hat{x}_i^i,\hat{x}_{-i}^i)\in\Omega^i$, with $\hat{x}_i^i\in\Omega_i$ as player $i$'s estimate of his action and $\hat{x}_{-i}^i\in\Omega_{-i}^{i}$ as the estimate of the other players whose actions interfere with player $i$'s cost function. Because of his imperfect available information, player $i$ uses $\hat{x}_{-i}^i(k)$  and updates his action as follows: if $i\in\{i_k,j_k\}$,
\begin{equation}\label{local_step}
x_i(k+1)=T_{\Omega_i}[x_i(k)-\alpha_{k,i} \nabla_{x_i}J_i(x_i(k),\hat{x}_{-i}^i(k))],
\end{equation}
otherwise, $x_i(k+1)=x_i(k)$. In \eqref{local_step}, $T_{\Omega_i}:\mathbb{R}\rightarrow\Omega_i$ is an Euclidean projection and $\alpha_{k,i}$ are diminishing step sizes such that
\begin{equation}
\label{diminish_step}
\sum_{k=1}^{\infty}\alpha_{k,i}^2<\infty,\qquad\sum_{k=1}^{\infty}\alpha_{k,i}=\infty\quad \forall i\in V.
\end{equation}
Note that $\alpha_{k,i}$ is inversely dependent on the number of updates $\nu_k(i)$ that each player $i$ has made until time $k$ (i.e., $\alpha_{k,i}=\frac{1}{\nu_k(i)}$). In \eqref{local_step}, players not involved in communication at $T(k)$ maintain their actions unchanged.
At this moment the updated actions are available for players to update their temporary estimates for every $i\in V,\,j\in\tilde{N}_I(i)$ as follows:
\begin{eqnarray}\label{temp_update}
\tilde{x}_j^i(k+1)=\begin{cases} \hat{x}_j^i(k),&\text{if }j\neq i \\ x_i(k+1),& \text{if }j=i.\end{cases}\end{eqnarray}
In \eqref{temp_update}, for $j\neq i$, player $i$'s estimate of player $j$'s action remains unchanged at the next iteration. However, for $j=i$, player $i$'s temporary estimate is updated by his action.

At this point, the players are ready to begin a new iteration from step 2. 
$\hfill\square$

In the following we write the algorithm in a compact form. Let $B=A+I_N\in \mathbb{R}^{N\times N}$, where $A=[a_{ij}]_{i,j\in V}$ is the adjacency matrix associated with $G_I$. Let also
\begin{equation}\label{s_ij}
s_{ij}:=\sum_{l=1}^{j}B(i,l)+\delta_{i\neq 1}\sum_{r=1}^{i-1}m_r,
\end{equation}
where
$
\delta_{i\neq 1} =
\begin{cases}
1,&\text{if } i\neq 1\\
0,&\text{if } i =1
\end{cases}
$.
Let  $e_{s_{ij}}$ be a unit vector in $\mathbb{R}^m$. For each pair $i,j\in V$, we assign a vector $E_j^i\in \mathbb{R}^m$,
\begin{equation}\label{E_j^i}
E_j^i=
\begin{cases}
e_{s_{ij}},&\text{if }i\in V,\,j\in\tilde{N}_I(i)\\
\textbf{0}_{m},&\text{if }i\in V,\,j\notin\tilde{N}_I(i).
\end{cases}
\end{equation}
The communication matrix $W(k)$ is defined as
\begin{eqnarray}\label{W_def}
W(k)\!:=\!I_m\!-\!\frac{1}{2}\sum_{l\in\text{ind}(i_k,j_k)}(E_l^{i_k}-E_l^{j_k})(E_l^{i_k}-E_l^{j_k})^T,
\end{eqnarray}
where $\text{ind}(i_k,j_k)\!:=\!\{d\!\in\! V\!:\!B(i_k,d)\!\cdot\! B(j_k,d)\!=\!1\!\}$ is the index set belong to $\tilde{N}_I(i_k)\cap\tilde{N}_I(j_k)$ for $i_k,\,j_k\in N_C$.
{\color{black}\begin{remark}
		Each player $i$ can only pass $\tilde{x}^i$ to his local neighbors. Since the dimension of the information that each player passes is different and dependent on the size of $N_I(i)$, we are unable to use the communication weight matrix
		for the fully coupled case $W(k)=\big(I_N-(e_{i_k}-e_{j_k})(e_{i_k}-e_{j_k})^T\big)\otimes I_N$ with $e_i\in \mathbb{R}^N$ as in \cite{salehisadaghiani2016distributed,nedic2011asynchronous}.
\end{remark}}
\begin{remark}\label{barabari_W_ha}
	$W(k)$ is a $(m \times m)$ generalized communication matrix, $N < m \leq N^2$.
	From \eqref{W_def} it follows  that $W(k)$ is a doubly stochastic matrix such that $W(k)^T\mathbf{1}_m=W(k)\mathbf{1}_m=\mathbf{1}_m$.
\end{remark}
\subsection{Example}
Consider a 4-player game in a network with the interference graph $G_I$ and the communication graph $G_C$ as in Fig.~2. Note that for $G_I$ in Fig.~2~(a), a selection of $G_m$ could be depicted as in Fig. 2 (c). One can verify that as  $G_m\subseteq G_C\subseteq G_I$, Assumption~\ref{connectivity} holds. In this setup $m_1=4$, $m_2=3$, $m_3=4$, $m_4=3$ and $m=\sum_{i=1}^{4}m_i=14$.
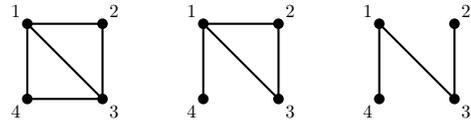
\begin{figure}[!htb]\label{ax_G_I}
	\centering
	\begin{tikzpicture}
	\fill (1/2,1/2) circle (2pt);
	\fill (-1/2,1/2) circle (2pt);
	\fill (-1/2,-1/2) circle (2pt);
	\fill (1/2,-1/2) circle (2pt);
	\draw[thick] (-1/2,1/2) node[above left, scale=0.7] {$1$} --
	(1/2,1/2) node[above right, scale=0.7] {$2$} --
	(1/2,-1/2) node[below right, scale=0.7] {$3$} --
	(-1/2,-1/2) node[below left, scale=0.7] {$4$};
	\draw[thick] (-1/2,-1/2) -- (-1/2,1/2);
	\draw[thick] (-1/2,1/2) -- (1/2,-1/2);
	\end{tikzpicture}
	\hspace{0.5cm}
	\begin{tikzpicture}
	\fill (1/2,1/2) circle (2pt);
	\fill (-1/2,1/2) circle (2pt);
	\fill (-1/2,-1/2) circle (2pt);
	\fill (1/2,-1/2) circle (2pt);
	\draw[thick] (-1/2,1/2) node[above left, scale=0.7] {$1$} --
	(1/2,1/2) node[above right, scale=0.7] {$2$} --
	(1/2,-1/2) node[below right, scale=0.7] {$3$};
	\draw[thick] (-1/2,-1/2) node[below left, scale=0.7] {$4$} -- (-1/2,1/2);
	\draw[thick] (-1/2,1/2) -- (1/2,-1/2);
	\end{tikzpicture}
	\hspace{0.5cm}
	\begin{tikzpicture}
	\fill (1/2,1/2) circle (2pt);
	\fill (-1/2,1/2) circle (2pt);
	\fill (-1/2,-1/2) circle (2pt);
	\fill (1/2,-1/2) circle (2pt);
	\draw[thick] (-1/2,1/2) node[above left, scale=0.7] {$1$}
	(1/2,1/2) node[above right, scale=0.7] {$2$} --
	(1/2,-1/2) node[below right, scale=0.7] {$3$};
	\draw[thick] (-1/2,-1/2) node[below left, scale=0.7] {$4$} -- (-1/2,1/2);
	\draw[thick] (-1/2,1/2) -- (1/2,-1/2);
	\end{tikzpicture}
	\caption{(a) Interference graph $G_I$ (b) Communication graph $G_C$ (c) A maximal triangle-free subgraph of $G_I$, $G_m$.}
\end{figure}

In the following tables we show the assignment of each vector $E_j^i\in \mathbb{R}^{14}$ to each $\tilde{x}_j^i\in\Omega_j\subset\mathbb{R}$.
\begin{center}
	\begin{tabular}{| l | c | r | r |}
		\hline
		$E_1^1=e_1$ & $E_2^1=e_2$ & $E_3^1=e_3$ & $E_4^1=e_4$ \\ \hline
		$E_1^2=e_5$ & $E_2^2=e_6$ & $E_3^2=e_7$ & $E_4^2=\textbf{0}_{14}$ \\ \hline
		$E_1^3=e_8$ & $E_2^3=e_9$ & $E_3^3=e_{10}$ & $E_4^3=e_{11}$ \\ \hline
		$E_1^4=e_{12}$ & $E_2^4=\textbf{0}_{14}$ & $E_3^4=e_{13}$ & $E_4^4=e_{14}$ \\
		\hline
	\end{tabular}
	\\$\downarrow$\\
	\begin{tabular}{| l | c | r | r |}
		\hline
		$\tilde{x}_1^1$ & $\tilde{x}_2^1$ & $\tilde{x}_3^1$ & $\tilde{x}_4^1$ \\ \hline
		$\tilde{x}_1^2$ & $\tilde{x}_2^2$ & $\tilde{x}_3^2$ & -- \\ \hline
		$\tilde{x}_1^3$ & $\tilde{x}_2^3$ & $\tilde{x}_3^3$ & $\tilde{x}_4^3$ \\ \hline
		$\tilde{x}_1^4$ & -- & $\tilde{x}_3^4$ & $\tilde{x}_4^4$ \\
		\hline
	\end{tabular}
\end{center}
Note that $e_i$ is a unit vector in $\mathbb{R}^{14}$ and $\textbf{0}_{14}$ is an $14\times 1$ vector whose entries are all 0.
Assume that players 2 and 3  communicate at $T(k)$, i.e., $i_k=2$, $j_k=3$. The set of indices in $\tilde{N}_I(2)\cap\tilde{N}_I(3)$ is denoted by $\text{ind}(2,3)=\{1,2,3\}$ and the communication matrix $W(k)$ is 
\begin{eqnarray}&&\hspace{-0.1cm}W(k)\!=\!I_{14}\!-\!\frac{1}{2}\big\{(e_5-e_{8})\!(e_5-e_{8})^T\!+\!(e_6-e_{9})\!(e_6-e_{9})^T\nonumber\\
&&\hspace{-0.1cm}+(e_7-e_{10})(e_7-e_{10})^T\big\}.\hspace{4.3cm}\square\nonumber\end{eqnarray}
Let $\bar{x}(k)$ be an intermediary variable. Let also $\tilde{x}(k):=\big[\tilde{x}^{1^T},\ldots,\tilde{x}^{N^T}\big]^T$ be the stack vector with the temporary estimates of all players and 
\begin{equation}\label{xbar}
\bar{x}(k)=W(k)\tilde{x}(k).
\end{equation} 
Then,
\begin{equation}\label{x_hat_tor}
\hat{x}_{-i}^i(k)=[\bar{x}_{r}(k)]_{r\in I(i)},
\end{equation}
where $I(i):=\{d:d=s_{ij},\ j\in N_I(i)\}$ and $s_{ij}$ as in \eqref{s_ij}.

{\color{black}The algorithm is as follows:
	\begin{algorithm}
		\caption{}
		\color{black}\begin{algorithmic}[1]
			\State \textbf{initialization} $\tilde{x}^i(0)\in\Omega^i\quad\forall i\in V$
			\For{$k=1,2,\ldots$ }
			\State \hspace{-0.5cm}$i_k\in V$ and $j_k\in N_C(i_k)$ communicate.
			\State \hspace{-0.5cm}$\bar{x}(k)=W(k)\tilde{x}(k)$, \eqref{xbar},\quad $\hat{x}^i_{-i}=[\bar{x}_r(k)]_{r\in I(i)}$, \eqref{x_hat_tor}.
			\State \hspace{-0.5cm}$x_{i}(k\!+\!1)\!=\!T_{\Omega_{i}}[x_{i}(k)\!-\!\alpha_{k,{i}} \nabla_{x_i}J_{i}(x_{i}(k)\!,\!\hat{x}_{-{i}}^{i}(k))]$
			if $i\in\{i_k,j_k\}$, \eqref{local_step}
			\Statex	$x_i(k+1)=x_i(k)$, otherwise.
			\State \hspace{-0.5cm}$\tilde{x}^i(k+1)=\hat{x}^i(k)+(x_i(k+1)-\hat{x}_i^i(k))e_i,\quad\forall i\in V$, \eqref{temp_update}.
			\EndFor
		\end{algorithmic}
	\end{algorithm}
}

\color{black}\section{Convergence for Diminishing Step Size}\label{convergence_diminish}
In this section we prove the convergence of the algorithm for diminishing step sizes.

{\color{black}The convergence proof has two parts:
\begin{enumerate}
	\item In Section~\ref{Conv_temp}, we prove almost sure convergence of the temporary estimate vector  $\tilde{x}(k)$  to an average consensus which is shown to be  $Z(k)$, the average of all temporary estimate vectors.
	\item In Section~\ref{conv_Nash}, we prove almost sure convergence of the players’ actions toward the Nash equilibrium. %
\end{enumerate}}
{\color{black}  
\subsection{Convergence of Temporary Estimates to An Average Consensus}\label{Conv_temp}
In this section, we define the average of all temporary estimates and prove that all the temporary estimates converge almost surely towards this point.}

Consider a memory $\mathcal{M}_k$ to denote the {\emph{sigma-field}} generated by the history up to time $k-1$ with $\mathcal{M}_0=\mathcal{M}_1=\{\tilde{x}^i(0),\text{ }i\in V\}$,
$$\mathcal{M}_k=\mathcal{M}_0\cup\Big\{(i_l,j_l); 1\leq l\leq k-1\Big\},\quad \forall k\geq 2.$$
For player $i$, let $m_i:=\text{deg}(i)+1$ where deg$(i)$ is the degree of vertex $i\in V$ in $G_I$. Let also $m:=\sum_{i=1}^{N}m_i$ and $\textbf{m}:=[m_1,\ldots,m_N]^T\in\mathbb{R}^N$.
\begin{remark}\label{mi>1}
	Assumption~\ref{connected_undirected} implies that $m_i>1$, $\forall i\in V$ and $m>N$. If the interference graph $G_I$ is a complete graph, i.e., fully coupled   cost functions, then $m=N^2$.
\end{remark}

Let $\tilde{x}(k)\in\mathbb{R}^m$ be the stack vector with temporary estimates of all players and $z(k)\in\mathbb{R}^N$ be the \emph{average} of all temporary estimate vectors,  $z(k):=\bar{H}\tilde{x}(k)$ where 
\begin{eqnarray}
&&\label{Hbar}\bar{H}:=\text{diag}(1./\textbf{m})H^T\in\mathbb{R}^{N\times m},\\
&&1./\textbf{m}:=[\frac{1}{m_1},\ldots,\frac{1}{m_N}]^T,\nonumber\\
&&\label{H}H:=[\sum_{i=1}^{N}E_1^i,\ldots,\sum_{i=1}^{N}E_N^i]\in\mathbb{R}^{m\times N}.
\end{eqnarray}
Let also $Z(k)\in\mathbb{R}^m$ denote the \emph{augmented average} of all temporary estimates, defined as follows:
\begin{equation}\label{ave_Z}
Z(k):=Hz(k)=H\bar{H}\tilde{x}(k)\in\mathbb{R}^{m}.
\end{equation}
We aim to prove almost sure convergence of $\tilde{x}(k)$ to $Z(k)$. Note that $H$ and $\bar{H}$ are non-square matrices. 
{\color{black}
\begin{remark}
Since every temporary estimate vector is not a full vector, the average of temporary estimates is not computed by $Z(k)=\frac{1}{N}(\textbf{1}_N^T\otimes I_N)\tilde{x}(k)$ as in \cite{salehisadaghiani2016distributed}. Rather, $H$ and $\bar{H}$ are defined to take element-wise average of the different number of temporary estimates associated with a specific player ($m_i=\text{deg}_{G_I}(i)+1$).
\end{remark}}
The convergence proof depends on some key properties of $W$ and $H$ given in Lemma~\ref{lemma_ext_stoch}-\ref{gamma_less_1}.
\begin{lemma}\label{lemma_ext_stoch}
Let W(k) and $H$ be defined in \eqref{W_def} and \eqref{H}. The following properties hold:
\begin{eqnarray}\label{ext_stoch}
&i)&\quad W^T(k)W(k)=W(k),\\
&ii)&\quad W(k)H=H,\label{W(k)H=Hlemma}\\
&iii)&\quad H^TW(k)=H^T.
\end{eqnarray}
\vspace{-1cm}
\end{lemma}
\par{\emph{Proof}}. See Appendix.
\begin{lemma}\label{QZ=0}
Let $Q(k):=W(k)-H\bar{H}W(k)$. Then $Q(k)Z(k)=\textbf{0}_m$.
\vspace{-0.25cm}
\end{lemma}
\par{\emph{Proof}}. See Appendix.
\begin{lemma}\label{R=1}
Let $R:=I_m-H\bar{H}$ where $H$, $\bar{H}$ defined in \eqref{H}, \eqref{Hbar}.  Then $\|R\|=1$, where the induced norm of $R$  is defined as  $\|R\|:=\sqrt{\lambda_{\text{max}}(R^TR)}$. 
\vspace{-0.25cm}
\end{lemma}
\par{\emph{Proof}}. See Appendix.\\
In the following, we define a parameter $\gamma$ which is related to $W(k)$ and plays an important role in the convergence proof of the players' actions to the Nash equilibrium, as well as in the convergence rate of the algorithm. Lemma~\ref{gamma_less_1} gives a strict upper bound on $\gamma$.
\begin{lemma}\label{gamma_less_1}
Let $Q(k):=W(k)-H\bar{H}W(k)$ and $\gamma=\lambda_{\max}\big(\mathbb{E}[Q(k)^TQ(k)]\big)$. Then $\gamma<1$.
\vspace{-0.25cm}
\end{lemma}
\par{\emph{Proof}}. See Appendix.
{\color{black}\begin{remark}
	Note that Lemma~2 in \cite{nedic2011asynchronous} cannot be used instead because $Q(k)$ is related to the matrices $H$ and $\bar{H}$ which are not vectors of all ones as in \cite{nedic2011asynchronous}.  
\end{remark}}
In the convergence proof we use the following lemma from \cite{polyak1987introduction} (Lemma~11, Chapter~2.2).
\begin{lemma}
\label{sigmond}
Let $V_k$, $u_k$, $\beta_k$ and $\zeta_k$ be non-negative random variables adapted to $\sigma$-algebra $\mathcal{M}_k$. If $\sum_{k=0}^{\infty}u_k<\infty$ {\color{black}a.s.}, $\sum_{k=0}^{\infty}\beta_k<\infty$ {\color{black}a.s.}, and $\mathbb{E}[V_{k+1}|\mathcal{M}_k]\leq(1+u_k)V_k-\zeta_k+\beta_k$ {\color{black}a.s.} for all $k\geq 0$, {\color{black}then $V_k$ converges a.s.} and $\sum_{k=0}^{\infty}\zeta_k<\infty$ {\color{black}a.s.}\vspace{-7pt}
\end{lemma}
Using Lemma~\ref{lemma_ext_stoch}-\ref{sigmond}, we show in the following that 
$\tilde{x}(k)$ converges to $Z(k)$.
\begin{theorem}\label{consensus1}
Let $\tilde{x}(k)$ be the stack vector with temporary estimates of all players and $Z(k)$ be its average as in \eqref{ave_Z}. Let also $\alpha_{k,\text{max}}=\max_{i\in V}\alpha_{k,i}$. Then under Assumptions~\ref{connected_undirected}-\ref{assump},
\begin{enumerate}[i)]
\item $\sum_{k=0}^{\infty}\alpha_{k,\text{max}}\|\tilde{x}(k)-Z(k)\|<\infty$\quad a.s.,
\item $\sum_{k=0}^{\infty}\|\tilde{x}(k)-Z(k)\|^2<\infty$\quad a.s.
\end{enumerate}
\end{theorem}
\par{\emph{Proof}}. See Appendix.\\
{\color{black}Note that by Theorem~\ref{consensus1}, $\|\tilde{x}(k)-Z(k)\|$ is almost surely square summable which implies the almost sure convergence of $\tilde{x}(k)$ to $Z(k)$.}

Theorem~\ref{consensus1} yields the following corollary for $x(k)$, represented as $x(k)=[\tilde{x}_1^1,\ldots,\tilde{x}_N^N]^T$.
\begin{corollary}\label{remark}
Let $z(k):=\bar{H}\tilde{x}(k)\in\mathbb{R}^N$ be the average of all players' temporary estimates. Under Assumptions~\ref{connected_undirected}-\ref{assump}, the following hold for players' actions $x(k)$:
\begin{enumerate}[i)]
\item $\sum_{k=0}^{\infty}\alpha_{k,\text{max}}\|x(k)-z(k)\|<\infty$\quad a.s.,
\item $\sum_{k=0}^{\infty}\|x(k)-z(k)\|^2<\infty$\quad a.s.
\end{enumerate}
\end{corollary}
{\color{black}\par{\emph{Proof}}. The proof follows by taking into account $x(k)=[\tilde{x}_i^i(k)]_{i\in V}$ and $Z(k)=Hz(k)$ \eqref{ave_Z}, and also using Theorem~\ref{consensus1}.
$\hfill\blacksquare$}

By Theorem~\ref{consensus1} and Corollary~\ref{remark}, the stack vector with temporary estimates of all players $\tilde{x}(k)$ converge to $Z(k)$ and all players' actions $x(k)$ converge to $z(k)$ as $k\rightarrow\infty$.
\begin{remark}
	Corollary~\ref{remark} implies that for  any $i\in V$ and  any $j\in\tilde{N}_I(i)$, $x_j$ converges toward $Z_{s_{ij}}(k)$  where $s_{ij}$ is as defined in \eqref{s_ij}.
\end{remark}

{\color{black}The next result shows almost sure convergence of $\bar{x}(k)$ toward $Z(k)$.
\begin{lemma}\label{x_bar_converge}
	Let $\tilde{x}(k)$ and $Z(k)$ be as in Theorem~\ref{consensus1}. Then for $\bar{x}(k)$, \eqref{xbar}, the~following~holds~under~Assumptions~\ref{connected_undirected}-\ref{assump},\vspace{-7pt}
	\begin{equation}
	\sum_{k=0}^{\infty}\mathbb{E}\Big[\|\bar{x}(k)-Z(k)\|^2\Big |\mathcal{M}_k\Big]<\infty\quad a.s.
	\end{equation}
\end{lemma}
\par{\emph{Proof}}. See Appendix.}\\
{\color{black}\subsection{Convergence of Players Actions to A Nash Equilibrium}\label{conv_Nash}
In this section, we use the result of Section~\ref{Conv_temp} to prove that all the players' actions converge towards a Nash equilibrium of the game.}

Let $p_i$ be the probability with which player $i$ updates his temporary estimate. The following theorem shows convergence to a Nash equilibrium of $\mathcal{G}$.
\begin{theorem}
\label{Prop_algo_1}
Let $x(k)$ and $x^*$ be all players' actions and the Nash equilibrium of $\mathcal{G}$, respectively. Under Assumptions~\ref{connected_undirected}-\ref{Lip_assump2}, the sequence $\{x(k)\}$ generated by the algorithm converges to $x^*$, almost surely.
\vspace{-0.25cm}
\end{theorem}
\par{\emph{Proof}}.\\
{\color{black}\emph{\textbf{Procedure:} First, we find an upper bound for $\mathbb{E}\Big[\|x(k+1)-x^*\|^2\Big |\mathcal{M}_k\Big]$ and simplify it to a similar format as in Lemma~\ref{sigmond}. Then we apply Lemma~\ref{sigmond} after verifying the conditions step by step and show that $\|x(k+1)-x^*\|^2$ converges almost surely to a non-negative limit point. Finally, we use $\sum_{k=0}^{\infty}\zeta_k<\infty$ as a result of Lemma~\ref{sigmond} and the strict monotonicity of $F$ to show that the limit point is $0$.}}

Firstly, using \eqref{local_step}, \eqref{NE_char} and the projection's non-expansive property, yields for $i\in\{i_k,j_k\}$, 
\begin{eqnarray}\label{Nash_eq_1}
\hspace{-0.20cm}&&\|x_i(k+1)-x_i^*\|^2\leq\|x_i(k)-x_i^*\|^2\nonumber\\
\hspace{-0.20cm}&&+\alpha_{k,i}^2\Big\|\nabla_{x_i}J_i(x_i(k),\hat{x}_{-i}^i(k))-\nabla_{x_i}J_i(x_i^*,{x_{-i}^{i\,*}})\Big\|^2\\
\hspace{-0.20cm}&&-2\alpha_{k,i}\Big(\!\nabla\!_{x_i}\!J_i(x_i(k),\hat{x}_{-i}^i(k))\!-\!\nabla\!_{x_i}\!J_i(x_i^*,{x_{-i}^{i\,*}})\Big)^T(x_i(k)\!-\!x_i^*).\nonumber
\end{eqnarray}
Adding and subtracting $\nabla_{x_i}J_i(x_i(k), [Z_t(k)]_{t\in I(i)})$  and  $\nabla_{x_i}J_i(x_i(k),x_{-i}^i(k))$ from the inner product term, using   \eqref{bounded} and $\pm2a^Tb\leq\|a\|^2+\|b\|^2$, yields for $i\in\{i_k,j_k\}$,
\begin{eqnarray}\label{sim5_Nash_eq}
\hspace{-0.20cm}&&\|x_i(k+1)-x_i^*\|^2\leq(1+2\alpha_{k,i}^2)\|x_i(k)-x_i^*\|^2+4C^2\alpha_{k,i}^2\nonumber\\
\hspace{-0.20cm}&&+\Big\|\nabla_{x_i}J_i(x_i(k),\hat{x}_{-i}^i(k))-\nabla_{x_i}J_i(x_i(k),[Z_t(k)]_{t\in I(i)})\Big\|^2\nonumber\\
\hspace{-0.20cm}&&+\Big\|\nabla\!_{x_i}\!J_i(x_i(k),[Z_t(k)]_{t\in I(i)})\!-\!\nabla\!_{x_i}\!J_i(x_i(k),x_{-i}^i(k))\Big\|^2\\
\hspace{-0.20cm}&&-2\alpha_{k,i}\Big(\!\nabla\!_{x_i}\!J_i(x_i(k),x_{-i}^i(k))\!-\!\nabla\!_{x_i}\!J_i(x_i^*,{x_{-i}^{i\,*}})\Big)^T\!(x_i(k)\!-\!x_i^*),\nonumber
\end{eqnarray}
where $I(i)$ is as defined in \eqref{x_hat_tor}.
For $i\notin\{i_k,j_k\}$, $x_i(k+1)=x_i(k)$ and $\|x_i(k+1)-x_i^*\|^2=\|x_i(k)-x_i^*\|^2$. One can combine these two cases together noting that for all $i\in V$, player $i$ updates his action with a given probability $p_i$. {\color{black}After taking the expected value, we obtain for $i\in V$:
\begin{eqnarray}\label{sim_prob1_Nash_eq}
&&\mathbb{E}\Big[\|x_i(k+1)-x_i^*\|^2\Big |\mathcal{M}_k\Big] \leq(1+2p_i\alpha_{k,i}^2)\|x_i(k)-x_i^*\|^2\nonumber\\
&&+4C^2p_i\alpha_{k,i}^2+p_i\mathbb{E}\Big[\Big\|\!\nabla\!_{x_i}\!J_i(x_i(k),\hat{x}_{-i}^i(k))\!\\
&&-\!\nabla\!_{x_i}\!J_i(x_i(k),[Z_t(k)]_{t\in I(i)})\Big\|^2\!\Big |\mathcal{M}_k\Big]\nonumber\\
&&+p_i\Big\|\nabla_{x_i}J_i(x_i(k),[Z_t(k)]_{t\in I(i)})-\nabla_{x_i}J_i(x_i(k),x_{-i}^i(k))\Big\|^2\!\nonumber\\
&&-2p_i\alpha_{k,i}\Big(\nabla_{x_i}J_i(\!x_i(k),x_{-i}^i(k))\nonumber\\
&&-\nabla_{x_i}J_i(x_i^*,x_{-i}^{i\,*})\Big)^T(x_i(k)-x_i^*).\nonumber
\end{eqnarray}
Let $p_{\text{max}}=\max_{i\in V}{p_i}$ and $p_{\text{min}}=\min_{i\in V}{p_i}$. Let also $\alpha_{k,\text{min}}=\min_{i\in V}{\alpha_{k,i}}$. After a simplification step (see the proof of Theorem~2 in \cite{salehisadaghiani2016distributed}) and summing over $i\in V$, for large enough $k$, we obtain the following using Assumption~\ref{Lip_assump2},
\begin{eqnarray*}\label{sim_prob33_Nash_eq}
	&&\mathbb{E}\Big[\|x(k+1)-x^*\|^2\Big |\mathcal{M}_k\Big]\leq\nonumber\\
	&&(1+2p_{\text{max}}\alpha_{k,\text{max}}^2+\frac{2p_\text{max}}{k^{3/2-q}p_{\text{min}}^2}+\frac{2p_\text{max}\rho^2}{k^{3/2-q}p_{\text{min}}^2}).\nonumber\\
	&&.\|x(k)-x^*\|^2+4NC^2p_{\text{max}}\alpha_{\text{max}}^2\nonumber\\
	&&+p_{\text{max}}L^2\sum_{i\in V}\mathbb{E}\Big[\|\hat{x}_{-i}^i(k)-[Z_t(k)]_{t\in I(i)}\|^2\Big |\mathcal{M}_k\Big]\nonumber\\
	&&+p_{\text{max}}L^2
	.\sum_{i\in V}\|[Z_t(k)]_{t\in I(i)}-x_{-i}^i(k)\|^2\nonumber\\
	&&-\frac{2}{k}(F(x(k))\!-\!F(x^*))^T(x(k)\!-\!x^*).
\end{eqnarray*}
We then apply Lemma~\ref{sigmond} for
\begin{eqnarray}
&&V_{k}:=\|x(k)-x^*\|^2,\nonumber\\
&& u_k:=2p_{\text{max}}\alpha_{k,\text{max}}^2+\frac{2p_\text{max}}{k^{3/2-q}p_{\text{min}}^2}+\frac{2p_\text{max}\rho^2}{k^{3/2-q}p_{\text{min}}^2},\nonumber\\
&&\beta_k:=p_{\text{max}}L^2\Big(\sum_{i\in V}\mathbb{E}\Big[\Big\|\hat{x}_{-i}^i(k)-[Z_t(k)]_{t\in I(i)}\Big\|^2\Big |\mathcal{M}_k\Big]\nonumber\\
&&
+\sum_{i\in V}\Big\|[Z_t(k)]_{t\in I(i)}-x_{-i}^i(k)\Big\|^2\Big)
+4NC^2p_{\text{max}}\alpha_{k,\text{max}}^2,\nonumber\\
&&\zeta_k:=\frac{2}{k}\Big(F(x(k))-F(x^*)\Big)^T(x(k)-x^*).\nonumber
\end{eqnarray}
By \eqref{diminish_step}, $\sum_{k=0}^{\infty}u_k<\infty$, and also by $\bar{x}(k)=W(k)\tilde{x}(k)$ \eqref{xbar}, Lemma~\ref{x_bar_converge} and Corollary~\ref{remark}, $\sum_{k=0}^{\infty}\beta_k<\infty$ a.s. Then by Lemma~\ref{sigmond}, $V_k$ converges almost surely to some positive limit and also $\sum_{k=0}^{\infty}\zeta_k<\infty$ a.s., hence 
\begin{enumerate}
	\item $\|x(k)-x^*\|^2$ converges almost surely,
	\item $\sum_{k=0}^{\infty}\frac{2}{k}\Big(F(x(k))-F(x^*)\Big)^T(x(k)-x^*)<\infty$ a.s.
\end{enumerate}
To complete the proof it only remains to show that $\|x(k)-x^*\|\rightarrow0$ a.s. This follows by the compactness of $\Omega$ (Assumption~\ref{assump}) and the strict monotonicity of $F$.
$\hfill\blacksquare$}
\color{black}\section{Non-Model-Based Approach for Convergence to a Nash Equilibrium}\label{Gradient_approximation}

We generalize the algorithm to the case when the players are not aware of their own cost functions (models) but use the measured values of their own realized costs at certain points. We employ the finite difference approximation method of the gradient (see \cite{polyak1987introduction}, Chapter~3.4).

Let $\tilde{\nabla}_{x_i}J_i(x_i(k),x_{-i}^i(k))$ denote the symmetric approximation of $\nabla_{x_i}J_i(x_i(k),x_{-i}^i(k))$ as the following:
\begin{eqnarray}
&&\tilde{\nabla}_{x_i}J_i(x_i(k),x_{-i}^i(k)):=\nonumber\\
&&\frac{J_i\big(x_i(k)+c_{k,i},x_{-i}^i(k)\big)-J_i\big(x_i(k)-c_{k,i},x_{-i}^i(k)\big)}{2c_{k,i}},\nonumber
\end{eqnarray}
where $c_{k,i}>0$ is a scalar perturbation on $x_i(k)$. By Taylor expansion, for a smooth cost function $J_i(\cdot)$, we obtain $J_i(x_i(k)\pm c_{k,i},x_{-i}^i(k))=J_i(x_i(k),x_{-i}^i(k))\pm c_{k,i}\nabla_{x_i}J_i(x_i(k),x_{-i}^i(k))+\frac{1}{2} c_{k,i}^2 \nabla_{x_ix_i}^2$ $ J_i(x_i(k),x_{-i}^i(k))\pm O(c_{k,i}^3)$. This implies that $\|\tilde{\nabla}_{x_i}J_i(x_i(k),x_{-i}^i(k))-\nabla_{x_i}J_i(x_i(k),x_{-i}^i(k))\|=O(c_{k,i}^2)$. However, smoothness of cost functions is a stringent assumption on this problem. A relaxation of this assumption is provided in the following:
\begin{assumption}\label{Assumption_finite_difference}
For every $i\in V$, $J_i(x_i(k),\cdot)$ is twice differentiable function in $x_i(k)$ and $\nabla_{x_ix_i}^2J_i(x_i(k),\cdot)$ satisfies a Lipschitz condition with constant $\eta>0$ in a $c_{k,i}$-ball around $x_i(k)$.
\end{assumption}
As in Lemma~1 in \cite{polyak1987introduction}, Chapter~3.4, one can show that,
\begin{equation}\label{Lemma_differnce_bounded}
\|\tilde{\nabla}_{x_i}J_i(x_i(k),\cdot)-\nabla_{x_i}J_i(x_i(k),\cdot)\|\leq\frac{\eta}{6}c_{k,i}^2.
\end{equation}
Note also that, by Assumption~\ref{assump} for $i\in V$ and
$x^i\in\Omega^i$ we have,
\begin{equation}\label{bounded_approx}
\|\tilde{\nabla}_{x_i}J_i(x^i)\|\leq H,\quad\text{for some }H>0.
\end{equation}
\begin{theorem}\label{Prop_algo_1_Approx}
Let $x(k)$ and $x^*$ be as in Theorem~\ref{Prop_algo_1}. Let also $\alpha_{k,i}$ be as in \eqref{diminish_step} and $c_{k,i}$ be a positive scalar such that,
\begin{equation}\label{alpha_c^2<infty}
\sum_{k=1}^{\infty}\alpha_{k,i}c_{k,i}^2<\infty,\quad \forall i\in V.
\end{equation}
Under Assumptions~\ref{connected_undirected}-\ref{Assumption_finite_difference}, the sequence $\{x(k)\}$, which is generated by the algorithm with the approximation of the gradient $\tilde{\nabla}_{x_i}J_i(.)$, converges to $x^*$, almost surely.
\end{theorem}
\par{\emph{Proof}}. As in \eqref{Nash_eq_1}, we find an upperbound for $\|x(k+1)-x^*\|^2$ considering that the gradient is approximated by $\tilde{\nabla}_{x_i}J_i(.)$ and then we use Lemma~\ref{sigmond} to prove the convergence.
\begin{eqnarray}\label{Nash_eq_1_approx}
\hspace{-0.2cm}&&\|x_i(k+1)-x_i^*\|^2\leq\|x_i(k)-x_i^*\|^2\nonumber\\
\hspace{-0.2cm}&&+\alpha_{k,i}^2\Big\|\tilde{\nabla}_{x_i}J_i(x_i(k),\hat{x}_{-i}^i(k))-\tilde{\nabla}_{x_i}J_i(x_i^*,{x_{-i}^{i\,*}})\Big\|^2\\
\hspace{-0.2cm}&&-2\alpha_{k,i}\Big(\!\tilde{\nabla}\!_{x_i}\!J_i(x_i(k),\hat{x}_{-i}^i(k))\!-\!\tilde{\nabla}\!_{x_i}\!J_i(x_i^*,{x_{-i}^{i\,*}})\Big)^T(x_i(k)\!-\!x_i^*).\nonumber
\end{eqnarray}
Using \eqref{bounded_approx} to estimate the 2nd term in the RHS of \eqref{Nash_eq_1_approx} and also adding and subtracting $\nabla_{x_i}J_i(x_i(k),\hat{x}_{-i}^i(k))$, $\nabla_{x_i}J_i(x_i(k), [Z_t(k)]_{t\in I(i)})$, $\nabla_{x_i}J_i(x_i(k),x_{-i}^i(k))$ and $\nabla_{x_i}J_i(x_i^*,{x_{-i}^{i*}})$ from the inner product term and using $\pm2a^Tb\leq\|a\|^2+\|b\|^2$, yields for $i\in\{i_k,j_k\}$,
\begin{eqnarray}\label{sim54_Nash_eq}
\hspace{-0.2cm}&&\|x_i(k+1)-x_i^*\|^2\leq(1+2\alpha_{k,i}^2)\|x_i(k)-x_i^*\|^2+4H^2\alpha_{k,i}^2\nonumber\\
\hspace{-0.2cm}&&+4\alpha_{k,i}x_{\max}\Big\|\tilde{\nabla}_{x_i}J_i(x_i(k),\hat{x}_{-i}^i(k))-\nabla_{x_i}J_i(x_i(k),\hat{x}_{-i}^i(k))\Big\|\nonumber\\
\hspace{-0.2cm}&&+\Big\|\nabla_{x_i}J_i(x_i(k),\hat{x}_{-i}^i(k))-\nabla_{x_i}J_i(x_i(k),[Z_t(k)]_{t\in I(i)})\Big\|^2\nonumber\\
\hspace{-0.2cm}&&+\Big\|\nabla\!_{x_i}\!J_i(x_i(k),[Z_t(k)]_{t\in I(i)})\!-\!\nabla\!_{x_i}\!J_i(x_i(k),x_{-i}^i(k))\Big\|^2\\
\hspace{-0.2cm}&&+4\alpha_{k,i}x_{\max}\Big\|\nabla\!_{x_i}\!J_i(x_i(k),x_{-i}^{i}(k))-\tilde{\nabla}\!_{x_i}\!J_i(x_i(k),x_{-i}^{i}(k))\Big\|\nonumber\\
\hspace{-0.2cm}&&-2\alpha_{k,i}\Big(\!\nabla\!_{x_i}\!J_i(x_i(k),x_{-i}^i(k))\!-\!\nabla\!_{x_i}\!J_i(x_i^*,{x_{-i}^{i\,*}})\Big)^T\!(x_i(k)\!-\!x_i^*),\nonumber
\end{eqnarray}
where $I(i)$ is as defined in \eqref{x_hat_tor}. In \eqref{sim54_Nash_eq}, we used the compactness of $\Omega_i$ (Assumption~3) which implies,\vspace{-0.3cm}
\begin{equation}\label{x_i_bound}
\|x_i(k)\|\leq x_{\max}.
\end{equation}
Using \eqref{Lemma_differnce_bounded}, yields for $i\in\{i_k,j_k\}$,
\begin{eqnarray}\label{sim55_Nash_eq}
&&\|x_i(k+1)-x_i^*\|^2\leq(1+2\alpha_{k,i}^2)\|x_i(k)-x_i^*\|^2\nonumber\\
&&+4H^2\alpha_{k,i}^2+\frac{4\eta x_{\max}}{3}\alpha_{k,i}c_{k,i}^2\\
&&+\Big\|\nabla_{x_i}J_i(x_i(k),\hat{x}_{-i}^i(k))-\nabla_{x_i}J_i(x_i(k),[Z_t(k)]_{t\in I(i)})\Big\|^2\nonumber\\
&&+\Big\|\nabla\!_{x_i}\!J_i(x_i(k),[Z_t(k)]_{t\in I(i)})\!-\!\nabla\!_{x_i}\!J_i(x_i(k),x_{-i}^i(k))\Big\|^2\nonumber\\
&&-2\alpha_{k,i}\Big(\!\nabla\!_{x_i}\!J_i(x_i(k),x_{-i}^i(k))\!-\!\nabla\!_{x_i}\!J_i(x_i^*,{x_{-i}^{i\,*}})\Big)^T\!(x_i(k)\!-\!x_i^*),\nonumber
\end{eqnarray}
The rest of the proof is similar to that of Theorem~\ref{Prop_algo_1}, noting that by \eqref{alpha_c^2<infty} $\sum_{k=1}^{\infty}\frac{4\eta x_{\max}}{3}\alpha_{k,i}c_{k,i}^2<\infty,\quad \forall i\in V$.$\hfill\blacksquare$
\color{black}\section{Convergence Rate}\label{convergence_rate}

In this section we compare the convergence rate of the algorithm proposed in Section~\ref{asynch} (denoted as Algorithm~1) with the algorithm in \cite{Salehisadaghiani2014Nash} (denoted as Algorithm~2). Algorithm~1 is an extension of Algorithm~2 which considers partially-coupled cost functions via an interference graph $G_I$. Algorithm~2 operates as if all cost functions are fully-coupled, or the interference graph is a complete graph $G$.
By Assumption~\ref{connectivity}, any feasible communication graph for Algorithm~1 has a lower bound $G_m$, however, the communication graph for Algorithm~2 can be any minimally connected subgraph of $G$, denoted as $G_{\min}$. Thus, since $G_{\min}\subseteq G_m$, we expect more iterations for Algorithm~1 than Algorithm~2 for the corresponding communication graphs from the point of view of parameters associated with $G_C$.
\begin{table*}[t]
\centering
\begin{tabular}{|c|c|c|}
\hline
& Algorithm~1 & Algorithm~2 [\cite{Salehisadaghiani2014Nash}]\\\hline
$G_C$ & $G_m\subseteq G_C^1\subseteq G_I\neq G$ & $G_{\min}\subseteq G_C^2\subseteq G$\\\hline
$\text{Deg}(i)+1$ & \multirow{2}{*}{$m_i$} & \multirow{2}{*}{$N$}\\
in $G_I$ \& $G$ & & \\\hline
 $\sum\limits_{i} (\text{Deg}(i)+1)$ & \multirow{2}{*}{$\sum_{i\in V} m_i=m$} & \multirow{2}{*}{$\sum_{i\in V} N=N^2$}\\
 in $G_I$ \& $G$ & & \\\hline
 $B$ & $B_1:=A_{G_I}+I_N$ & $B_2:=A_{G}+I_N=\textbf{1}_N\textbf{1}_N^T$ \\\hline
 $E_j^i\in \mathbb{R}^m$ \eqref{E_j^i} & {$\!E_j^{i,1}=\begin{cases}e_{s_{ij}},& \text{if }i\in V,\,j\in\tilde{N}_I(i)\\
\textbf{0}_{m\times 1},& \text{if }i\in V,\,j\notin\tilde{N}_I(i)\end{cases}$} & $E_j^{i,2}=e_{(i-1)N+j}$, $i,j\in V$ \\\hline
\multirow{2}{*}{$W(k)$ \eqref{W_def}} & \multicolumn{2}{c|}{$\displaystyle W_1(k):=I_m-\frac{1}{2}\sum_{l\in\text{ind}(i_k,j_k)}(E_l^{i_k,1}-E_l^{j_k,1})(E_l^{i_k,1}-E_l^{j_k,1})^T$} \\
 & \multicolumn{2}{c|}{$\displaystyle W_2(k):=I_{N^2}-\frac{1}{2}\sum_{l\in V}(e_{(i_k-1)N+l}-e_{(j_k-1)N+l})(e_{(i_k-1)N+l}-e_{(j_k-1)N+l})^T$} \\\hline
$H$ \eqref{H} & $H_1:=[\sum_{i=1}^{N}E_1^{i,1},\ldots,\sum_{i=1}^{N}E_N^{i,1}]\in\mathbb{R}^{m\times N}$ & $H_2:=I_N\otimes\textbf{1}_N\in\mathbb{R}^{N^2\times N}$\\\hline
$\bar{H}$ \eqref{Hbar} & $\bar{H}_1:=\text{diag}(1./\textbf{m})H_1^T\in\mathbb{R}^{N\times m}$ & $\bar{H}_2:=\frac{1}{N}H_2^T=\frac{1}{N}(I_N\otimes\textbf{1}_N^T)\in\mathbb{R}^{N\times N^2}$\\\hline
$Q(k)$ & \multicolumn{2}{c|}{$Q_1(k):=W_1(k)-H_1\bar{H}_1W_1(k)=W_1(k)-H_1\text{diag}(1./\textbf{m})H_1^T$}\\
(Lemma~\ref{QZ=0}) & \multicolumn{2}{c|}{$Q_2(k):=W_2(k)-\frac{1}{N}H_2H_2^T=W_2(k)-\frac{1}{N}(I_N\otimes\textbf{1}_N\textbf{1}_N^T)$}  \\\hline
$R$ (Lemma~\ref{R=1}) & $R_1:=I_m-H_1\bar{H}_1=I_m-H_1\text{diag}(1./\textbf{m})H_1^T$ & $R_2:=I_{N^2}-\frac{1}{N}H_2H_2^T=I_N\otimes(I_N-\frac{1}{N}\textbf{1}_N\textbf{1}_N^T)$\\\hline
\end{tabular}\\\vspace{0.5cm}
Table I
\end{table*}
In TABLE~I we summarize the differences between the parameters of Algorithms~1 and 2. Note that we distinguish between the parameters for Algorithms 1, 2 by using subscripts 1, 2 (superscripts 1, 2 for $G_C$, $N_C$ and $E_j^i$). To avoid any confusion note also that $\lambda_2$ denotes the second largest eigenvalue and its index has nothing to do with the subscript associated with Algorithm~2.

We compare Algorithms~1 and 2 relative to the interference graph $G_I$. We can show that for each iteration, Algorithm~1 takes less time than Algorithm~2 since less information (fewer estimates) is needed to be exchanged. For the sake of comparison, we assume that both algorithms run over the same $G_C\supseteq G_m$. Let $r$ be the time required to exchange an estimate, and let $s$ be the time required to process a full gradient. Note that the processing time for the gradient is linearly dependent on the data set. We ignore the time required to compute the projection in the local step. Thus for each iteration, the average time required to exchange all the estimates between players and to update the actions under Algorithm~1 is
\begin{equation}\label{average_time_interference}
T_\text{av}^1:=\sum_{i\in V}\sum_{j\in N_C(i)}\frac{1}{N}p_{ij}\Big(|N_I(i)\cap \tilde{N}_I(j)|r+\frac{m_i}{N}s\Big),
\end{equation}
where $p_{ij}$ is the probability that players $i$ and $j$ contact each other. $|N_I(i)\cap \tilde{N}_I(j)|r$ is the time required for player $i$ to obtain all the necessary estimates of player $j$. $\frac{m_i}{N}s$ is the time required to compute $\nabla_{x_i}J_i(x_i,x_{-i}^i)$, noting that $s$ is the processing time for computing the full gradient $\nabla_{x_i}J_i(x_i,x_{-i})$.
In Algorithm~2 the average time for each iteration is computed by replacing $|N_I(i)\cap \tilde{N}_I(j)|$ and $m_i$ in \eqref{average_time_interference} with $N-1$ and $N$, respectively. Then,
we obtain,
\begin{equation}\label{average_time_communication}
T_\text{av}^2:=\sum_{i\in V}\sum_{j\in N_C(i)}\frac{1}{N}p_{ij}\Big((N-1)r+s\Big),
\end{equation}
Note that $|N_I(i)\cap N_I(j)|\leq N-1$ and $m_i\leq N$ which implies $T_\text{av}^1\leq T_\text{av}^2$.\\
Next, we discuss the convergence time (in number of iterations)  required for each algorithm. To simplify the analysis, we assume constant step sizes (i.e., $\alpha_{k,i}=\alpha_i$). Note that for constant step sizes there exists a steady-state offset between $x(k)$ and the Nash equilibrium $x^*$, see \cite{salehisadaghiani2016distributed}. Let this minimum value of error be denoted by $d^*$, i.e., $\inf_{k}\|x(k)-x^*\|=d^*$. We use a modified \emph{$\epsilon$-averaging time} similar to Definition~1 in \cite{boyd2006randomized} for the convergence time.
\begin{definition}\label{epsilon_averaging}
For any $0<\epsilon<1$, the $\epsilon$-averaging time of an algorithm, $N_\text{av}(\epsilon)$,  is defined as
\begin{equation}
\hspace{-0.115cm}N_\text{av}(\epsilon)\!:=\!\sup_{x(0)}\!\inf\!\Big\{k\!:\!\text{Pr}\Big(\frac{\|x(k)\!-\!x^*\|\!-\!d^*}{\|x(0)\|}\!\geq\!\epsilon\Big)\!\leq\!\epsilon\Big\}.
\end{equation}
\vspace{-0.75cm}
\end{definition}
By Definition~\ref{epsilon_averaging}, $N_\text{av}(\epsilon)$ is the minimum number of iterations it takes for $\|x(k)-x^*\|$ to approach an $\epsilon$-ball around $d^*$ with a high probability, regardless of the initial condition $x(0)$. The following assumption guarantees $N_\text{av}(\epsilon)$ to be well-defined.
\begin{assumption}\label{non_zero_minimum_value}
We assume a non-zero minimum value, denoted by $x_\text{min}(0)\neq 0$, for the norm of the initial action of player $i$ for $i\in V$, i.e., $\|x_i(0)\|\geq x_\text{min}(0)>0$.
\end{assumption}
We obtain a lower bound for the $\epsilon$-averaging time under Algorithms~1 and~2 by applying \emph{Markov's inequality}: for any non-negative random variable $X$ and $\epsilon>0$, the following holds:
\begin{equation}\label{markov_inequality}
\text{Pr}(X\geq\epsilon)\leq\frac{\mathbb{E}[X]}{\epsilon}.
\end{equation}
For constant step sizes we consider the following assumption rather than Assumption~\ref{Lip_assump}.
\begin{assumption}\label{strongly_monotone}
$F:\Omega\rightarrow\mathbb{R}^N$ is strongly monotone on $\Omega$ with a constant $\mu>0$, i.e.,
\begin{equation}
(F(x)-F(y))^T(x-y)\geq\mu\|x-y\|^2\quad\forall x,y\in\Omega.
\end{equation}
\end{assumption}
\begin{theorem}\label{theorem_convergence_rate}
Let $\alpha_i$ be constant step sizes which satisfy $0<\phi<1$ where,
\begin{equation}\label{condition}
\phi:=1+(1+\rho^2+2\alpha_\text{max})p_\text{max}\alpha_{\text{max}}-(1+\rho^2+2\mu)p_{\text{min}}\alpha_{\text{min}},
\end{equation}
with $p_{\text{max}}=\max_{i\in V}{p_i}$, $p_{\text{min}}=\min_{i\in V}{p_i}$, $\alpha_{\text{max}}=\max_{i\in V}{\alpha_{i}}$, $\alpha_{\text{min}}=\min_{i\in V}{\alpha_{i}}$, $\rho$ be the Lipschitz constant of $F$ and $\mu$ be the positive constant for the strong monotonicity property of $F$. Under Assumptions~\ref{connected_undirected}-\ref{assump}, \ref{Lip_assump2}, \ref{non_zero_minimum_value}, \ref{strongly_monotone}, the $\epsilon$-averaging time $N_{\text{av}}(\epsilon)$ has a lower bound as follows:
\begin{equation*}\label{N_av}
N_\text{av}(\epsilon)\geq\frac{\log \frac{a}{\epsilon^{3}-b}}{\log\frac{1}{\sqrt{\gamma}}},
\end{equation*}
where $\gamma=\lambda_{\max}\big(\mathbb{E}[Q(k)^TQ(k)]\big)$ (as in Lemma~\ref{gamma_less_1}), $Q(k):=W(k)-H\bar{H}W(k)$, and $a, b$ are positive and increasing with $\gamma$.
\end{theorem}
\par{\emph{Proof}}.\\
{\color{black}\emph{\textbf{Procedure:} The proof follows by bounding $\mathbb{E}\Big[\|x(k+1)-x^*\|^2\Big]$ using an upper bound for $\mathbb{E}\Big[\|\tilde{x}(k+1)-Z(k+1)\|\Big]$. Then we use Markov's inequality \eqref{markov_inequality} to obtain a lower bound for the $\epsilon$-averaging time $N_\text{av}(\epsilon)$.}}

First, we start to find an upper bound for $\mathbb{E}\Big[\|\tilde{x}(k+1)-Z(k+1)\|\Big]$. 
As in the proof of Theorem~1 (Equation~\eqref{term1&term2_karshode}), one can obtain,
\begin{eqnarray}\label{x_tildeZ}
&&\mathbb{E}\Big[\|\tilde{x}(k+1)-Z(k+1)\|\Big]\leq\sqrt{\gamma}\mathbb{E}\Big[\|\tilde{x}(k)-Z(k)\|\Big]\nonumber\\
&&+\frac{\sqrt{2}}{2}\!\sum_{i\in\{i_k,j_k\}}\!\mathbb{E}\Big[\|\tilde{x}_i^i(k)-\tilde{x}_i^j(k)\|\Big]\!+\!2\alpha_{\text{max}}C.
\end{eqnarray}
In  \eqref{x_tildeZ}, we upper bound $\mathbb{E}\Big[\|\tilde{x}_i^i(k)-\tilde{x}_i^j(k)\|\Big]$.
By \eqref{excluding}, \eqref{local_step}, \eqref{temp_update} and \eqref{x_i_bound} we obtain for $i,j\in\{i_k,j_k\}$,
\begin{eqnarray}\label{expo1}
&&\mathbb{E}\Big[\|\tilde{x}_i^i(k+1)-\tilde{x}_i^j(k+1)\|\Big]\nonumber\\
&&\leq\mathbb{E}\Big[\|\frac{\tilde{x}_i^i(k)-\tilde{x}_i^j(k)}{2}-\alpha_i\nabla_{x_i}J_i(x_i(k),\hat{x}_{-i}^i(k))\|\Big]\nonumber\\
&&\leq\frac{1}{2}\mathbb{E}\Big[\|\tilde{x}_i^i(k)-\tilde{x}_i^j(k)\|\Big]+\alpha_\text{max}C\nonumber\\
&&\leq(\frac{1}{2})^{k+1}\mathbb{E}\Big[\|\tilde{x}_i^i(0)-\tilde{x}_i^j(0)\|\Big]+\sum_{t=0}^{k}(\frac{1}{2})^t\alpha_\text{max}C\nonumber\\
&&\leq(\frac{1}{2})^{k}x_\text{max}+\sum_{t=0}^{k}(\frac{1}{2})^t\alpha_\text{max}C\leq C_1\sum_{t=0}^{k}(\frac{1}{2})^t\leq2C_1,
\end{eqnarray}
where $C_1:=\max\{\alpha_{\max}C,x_{\max}\}$. Substituting \eqref{expo1} into \eqref{x_tildeZ}, one can obtain,
\begin{eqnarray}\label{error_term1_2}
&&\mathbb{E}\Big[\|\tilde{x}(k+1)-Z(k+1)\|\Big]\leq\sqrt{\gamma}\mathbb{E}\Big[\|\tilde{x}(k)-Z(k)\|\Big]\nonumber\\
&&+\sqrt{2}\sum_{i\in\{i_k,j_k\}}C_1+\!2\alpha_{\text{max}}C\leq\sqrt{\gamma}^{k+1}\mathbb{E}\Big[\|\tilde{x}(0)-Z(0)\|\Big]\nonumber\\
&&+(2\sqrt{2}C_1+\!2\alpha_{\text{max}}C)\sum_{t=0}^k\sqrt{\gamma}^t
\leq C_2\sqrt{\gamma}^{k+1}+C_{21},
\end{eqnarray}
where $C_2:=\sqrt{N}x_{\max}$, $C_{21}:=\frac{2\sqrt{2}C_1+\!2\alpha_{\text{max}}C}{1-\sqrt{\gamma}}$ and  we used $\sqrt{N}x_{\min}\leq\|x\|\leq\sqrt{N}x_{\max}$,  by \eqref{x_i_bound}.
Taking \eqref{x_tildeZ} into account, one can upper bound the following squared-norm term:
\begin{eqnarray}\label{error_term_squared}
&&\mathbb{E}\Big[\|\tilde{x}(k+1)-Z(k+1)\|^2\Big]\leq\gamma\mathbb{E}\Big[\|\tilde{x}(k)-Z(k)\|^2\Big]\nonumber\\
&&+\frac{1}{2}\sum_{i\in\{i_k,j_k\}}\mathbb{E}\Big[\|\tilde{x}_i^i(k)-\tilde{x}_i^j(k)\|^2\Big]+4\alpha_{\text{max}}^2C^2\nonumber\\
&&+2\mathbb{E}\Big[\Big(\sqrt{\gamma}\|\tilde{x}(k)-Z(k)\|\Big)\nonumber\\
&&.\Big(\frac{\sqrt{2}}{2}\sum_{i\in\{i_k,j_k\}}\|\tilde{x}_i^i(k)-\tilde{x}_i^j(k)\|+2\alpha_{\text{max}}C\Big)\Big].
\end{eqnarray}
To simplify \eqref{error_term_squared}, we first deal with the second term and then with the last term of the RHS. By \eqref{expo1} we arrive at,
\begin{eqnarray}\label{expo2}
\hspace{-0.2cm}&&\mathbb{E}\Big[\|\tilde{x}_i^i(k+1)-\tilde{x}_i^j(k+1)\|^2\Big]\leq\frac{1}{4}\mathbb{E}\Big[\|\tilde{x}_i^i(k)-\tilde{x}_i^j(k)\|^2\Big]\\
\hspace{-0.2cm}&&+\alpha_{\text{max}}^2C^2+\alpha_{\text{max}}C\mathbb{E}\Big[\|\tilde{x}_i^i(k)-\tilde{x}_i^j(k)\|\Big]\leq C_3\sum_{t=0}^{k+1}(\frac{1}{4})^t\leq\frac{4}{3}C_3,\nonumber
\end{eqnarray}
where $C_3:=\max\{x_{\max}^2,\alpha_{\text{max}}^2C^2+2\alpha_{\text{max}}CC_1\}$. Multiplication of \eqref{x_tildeZ} and \eqref{expo1} yields, 
\begin{eqnarray}\label{multipli_bade_expec}
\hspace{-0.2cm}&&\mathbb{E}\Big[\Big(\frac{1}{2}\sum_{i\in\{i_k,j_k\}}\|\tilde{x}_i^i(k+1)-\tilde{x}_i^j(k+1)\|\Big)\nonumber\\
\hspace{-0.2cm}&&\Big(\|\tilde{x}(k+1)-Z(k+1)\|\Big)\Big]\nonumber\\
\hspace{-0.2cm}&&\leq\!\frac{\sqrt{\gamma}}{2}\!\mathbb{E}\Big[\Big(\!\frac{1}{2}\sum_{i\in\{\!i_k,j_k\!\}}\!\|\tilde{x}_i^i(k)-\tilde{x}_i^j(k)\|\!\Big)\|\tilde{x}(k)-Z(k)\|\Big]\nonumber\\
\hspace{-0.2cm}&&+\frac{\sqrt{2}}{4}\sum_{i\in\{\!i_k,j_k\!\}}\mathbb{E}\Big[\|\tilde{x}_i^i(k)-\tilde{x}_i^j(k)\|^2\Big] \nonumber\\
\hspace{-0.2cm}&&+\frac{1+\sqrt{2}}{2}\alpha_{\text{max}}C\sum_{i\in\{\!i_k,j_k\!\}}\!\mathbb{E}\Big[\|\tilde{x}_i^i(k)-\tilde{x}_i^j(k)\|\Big]\\
\hspace{-0.2cm}&&+\sqrt{\gamma}\alpha_{\text{max}}C\mathbb{E}\Big[\|\tilde{x}(k)-Z(k)\|\Big]\!+\!2\alpha_{\text{max}}^2C^2\leq C_4\sqrt{\gamma}^{k+1}+C_{41},\nonumber
\end{eqnarray}\vspace{-0.2cm}
where $C_{41}:=\frac{1}{{1-\frac{\sqrt{\gamma}}{2}}}\big(\frac{2\sqrt{2}}{3}C_3+2(1+\sqrt{2})\alpha_{\max}CC_1+CC_{21}\alpha_{\max}\sqrt{\gamma}+2\alpha_{\max}^2C^2\big)+Nx_{\max}^2\frac{\sqrt{\gamma}}{2}$ and $C_4:=\frac{CC_2\alpha_{\max}}{1-\frac{\sqrt{\gamma}}{2}}$.
Using  \eqref{expo2}, \eqref{multipli_bade_expec} and \eqref{error_term1_2} on the RHS of  \eqref{error_term_squared}, one can obtain,
\begin{eqnarray}\label{error_term_squared222}
&&\mathbb{E}\Big[\|\tilde{x}(k+1)-Z(k+1)\|^2\Big]\leq\nonumber\\
&&\leq\gamma\mathbb{E}\Big[\|\tilde{x}(k)-Z(k)\|^2\Big]+\frac{4}{3}C_3+4\alpha_{\max}^2C^2\nonumber\\
&&+2\sqrt{2\gamma}(C_4\sqrt{\gamma}^k+C_{41})+4\sqrt{\gamma}\alpha_{\max}C(C_2\sqrt{\gamma}^k+C_{21})\nonumber\\
&&\leq C_5\sqrt{\gamma}^{k+1}+C_{51},
\end{eqnarray}\vspace{-0.2cm}
where $C_{51}:=\frac{\frac{4}{3}C_3+4\alpha_{\max}^2C^2+(2\sqrt{2}C_{41}+4\alpha_{\max}CC_{21})\sqrt{\gamma}}{1-\gamma}+\gamma Nx_{\max}^2$ and $C_5:=\frac{2\sqrt{2}C_4+4\alpha_{\max}CC_2}{1-\gamma}$.
We put the last piece of the puzzle in place by finding an upper bound for $\mathbb{E}\Big[\|x(k+1)-x^*\|^2\Big]$. Using \eqref{local_step}, for $i\in \{i_k,j_k\}$ and  bringing in $Z(k)$ it follows that for $i\in\{i_k,j_k\}$,
\begin{eqnarray}\label{error_avali}
\hspace{-0.2cm}&&\|x_i(k+1)-x_i^*\|^2\leq\nonumber\\
\hspace{-0.2cm}&&\Big\|x_i(k)\!-\!x_i^*\!-\!\alpha_{i}\!\Big(\!\nabla\!_{x_i}\!J_i(x_i(k),\hat{x}_{-i}^i(k))\!-\!\nabla\!_{x_i}\!J_i(x_i^*,{x_{-i}^{i\,*}})\Big)\Big\|^2 \nonumber\\
\hspace{-0.2cm}&&\leq(1+2\alpha_{\text{max}}^2)\|x_i(k)-x_i^*\|^2+4C^2\alpha_{\text{max}}^2\nonumber\\
\hspace{-0.2cm}&&+L_i^2\Big\|\hat{x}_{-i}^i(k)-[Z_t(k)]_{t\in I(i)}\Big\|^2+L_i^2\Big\|x_{-i}^i(k)-[Z_t(k)]_{t\in I(i)}\Big\|^2\nonumber\\
\hspace{-0.2cm}&&-2\alpha_{i}\Big(\!\nabla\!_{x_i}\!J_i(x_i(k),x_{-i}^i(k))\!-\!\nabla\!_{x_i}\!J_i(x_i^*,{x_{-i}^{i\,*}})\Big)^T\!(x_i(k)\!-\!x_i^*),
\end{eqnarray}
where $I(i)$ is as defined in \eqref{x_hat_tor}. For $i\notin\{i_k,j_k\}$, $x_i(k+1)=x_i(k)$. Similar to the proof of Theorem~\ref{Prop_algo_1}, we combine these two cases for $i\in V$ and take conditional expected value to obtain, 
\begin{eqnarray}\label{error_bade_Lipschitz_expected}
&&\mathbb{E}\Big[\|x_i(k+1)-x_i^*\|^2\Big|\mathcal{M}_k\Big]\leq(1+2p_i\alpha_{\text{max}}^2)\|x_i(k)-x_i^*\|^2\nonumber\\
&&+4C^2p_i\alpha_{\text{max}}^2
+L_i^2p_i\mathbb{E}\Big[\Big\|\hat{x}_{-i}^i(k)-[Z_t(k)]_{t\in I(i)}\Big\|^2\Big|\mathcal{M}_k\Big]\nonumber\\
&&\vspace{-0.5cm}
+L_i^2p_i\Big\|x_{-i}^i(k)-[Z_t(k)]_{t\in I(i)}\Big\|^2\\
&&
-2p_i\alpha_{i}\Big(\!\nabla\!_{x_i}\!J_i(x_i(k),x_{-i}^i(k))\!-\!\nabla\!_{x_i}\!J_i(x_i^*,{x_{-i}^{i\,*}})\Big)^T\!(x_i(k)\!-\!x_i^*).\nonumber
\end{eqnarray}
Adding and subtracting $\alpha_{\min}p_{\min}$ from $\alpha_ip_i$ in the last term, we obtain after some manipulation,
\begin{eqnarray}\label{error_bade_Lipschitz_expected2}
&&\mathbb{E}\Big[\|x_i(k+1)-x_i^*\|^2\Big|\mathcal{M}_k\Big]\leq\nonumber\\
&&(1+p_{\max}\alpha_{\max}-p_{\min}\alpha_{\min}+2p_i\alpha_{\text{max}}^2)\|x_i(k)-x_i^*\|^2\nonumber\\
&&
+4C^2p_i\alpha_{\text{max}}^2
+L_i^2p_i\mathbb{E}\Big[\Big\|\hat{x}_{-i}^i(k)-[Z_t(k)]_{t\in I(i)}\Big\|^2\Big|\mathcal{M}_k\Big]\nonumber\\
&&
+L_i^2p_i\Big\|x_{-i}^i(k)-[Z_t(k)]_{t\in I(i)}\Big\|^2+(p_{\max}\alpha_{\max}-p_{\min}\alpha_{\min})\nonumber\\
&&
.\|\!\nabla\!_{x_i}\!J_i(x_i(k),x_{-i}^i(k))\!-\!\nabla\!_{x_i}\!J_i(x_i^*,{x_{-i}^{i\,*}})\|^2-2p_{\min}\alpha_{\min}\nonumber\\
&&
.\Big(\!\nabla\!_{x_i}\!J_i(x_i(k),x_{-i}^i(k))\!-\!\nabla\!_{x_i}\!J_i(x_i^*,{x_{-i}^{i\,*}})\Big)^T\!(x_i(k)\!-\!x_i^*).\nonumber
\end{eqnarray}
Summing the foregoing  over all $i\in V$, and using  $L=\max_{i\in V}{L_i}$ we arrive at, 
\begin{eqnarray}\label{error_bade_Lipschitz_expected3}
&&\mathbb{E}\Big[\|x(k+1)-x^*\|^2\Big|\mathcal{M}_k\Big]\leq\nonumber\\
&&
(1+p_{\max}\alpha_{\max}-p_{\min}\alpha_{\min}+2p_{\max}\alpha_{\text{max}}^2)\|x(k)-x^*\|^2\nonumber\\
&&+4NC^2p_{\max}\alpha_{\text{max}}^2\nonumber\\
&&
+L^2p_{\max}\sum_{i\in V}\mathbb{E}\Big[\Big\|\hat{x}_{-i}^i(k)-[Z_t(k)]_{t\in I(i)}\Big\|^2\Big|\mathcal{M}_k\Big]\nonumber\\
&&
+L^2p_{\max}\sum_{i\in V}\Big\|x_{-i}^i(k)-[Z_t(k)]_{t\in I(i)}\Big\|^2\nonumber\\
&&
+(p_{\max}\alpha_{\max}-p_{\min}\alpha_{\min})\|F(x(k))-F(x^*)\|^2\nonumber\\
&&
-2p_{\min}\alpha_{\min}\Big(F(x(k))-F(x^*)\Big)^T\!(x(k)\!-\!x^*).
\end{eqnarray}
Using Remark~\ref{Lipscitz_F} in the fifth term  and Assumption~\ref{strongly_monotone} in the last term and taking the expected value, 
yields,
\begin{eqnarray}\label{error_bade_Lipschitz_expected5}
&&\mathbb{E}\Big[\mathbb{E}\Big[\|x(k+1)-x^*\|^2\Big|\mathcal{M}_k\Big]\Big]=\mathbb{E}\Big[\|x(k+1)-x^*\|^2\Big]\nonumber\\
&&\leq\phi\mathbb{E}\Big[\|x(k)-x^*\|^2\Big]+4NC^2p_{\max}\alpha_{\text{max}}^2\nonumber\\
&&+L^2p_{\max}\sum_{i\in V}\mathbb{E}\Big[\Big\|\hat{x}_{-i}^i(k)-[Z_t(k)]_{t\in I(i)}\Big\|^2\Big]\nonumber\\
&&+L^2p_{\max}\sum_{i\in V}\mathbb{E}\Big[\Big\|x_{-i}^i(k)-[Z_t(k)]_{t\in I(i)}\Big\|^2\Big],
\end{eqnarray}
where $\phi$ is as in \eqref{condition}, $\rho>0$ and $\mu>0$ are the Lipschitz and the strong monotonicity constants, respectively. Using \eqref{error_term_squared222} into \eqref{error_bade_Lipschitz_expected5}, yields the following upper bound:
\begin{eqnarray}\label{error_bade_Lipschitz_expected6}
&&\mathbb{E}\Big[\|x(k+1)-x^*\|^2\Big]\leq\phi\mathbb{E}\Big[\|x(k)-x^*\|^2\Big]\nonumber\\
&&+4NC^2p_{\max}\alpha_{\text{max}}^2+2L^2p_{\max}(C_5\sqrt{\gamma}^k+C_{51})\nonumber\\
&&\leq C_6\Big(\frac{1-\phi^{k+2}}{1-\phi}\Big)+\sum_{t=0}^{k}\phi^t(2C_5L^2p_{\max}\sqrt{\gamma}^k)\nonumber\\
&&\leq C_7+C_8\sqrt{\gamma}^k,
\end{eqnarray}
where $C_7:=\frac{C_6}{1-\phi}$, $C_8:=\frac{2L^2p_{\max}C_5}{1-\phi}$ and $C_6:=\max\{Nx_{\max}^2,4NC^2p_{\max}\alpha_{\text{max}}^2+2L^2p_{\max}C_{51}\}$.
Recall that $\inf_{k}\|x(k)-x^*\|=d^*$. Then,
\begin{equation}\label{d*<C7}
{d^*}^2=\inf_{k}\|x(k)-x^*\|^2\leq\lim_{k\rightarrow\infty}\mathbb{E}[\|x(k)-x^*\|^2]\leq C_7.
\end{equation}
Since $0<\phi<1$, by using Markov's inequality \eqref{markov_inequality} and \eqref{error_bade_Lipschitz_expected6}  the following inequality follows,
\begin{eqnarray}\label{probability_inequality}
&&\text{Pr}\Big(\frac{\|x(k)-x^*\|-d^*}{\|x(0)\|}\geq\epsilon\Big)=\text{Pr}\Big(\frac{(\|x(k)-x^*\|-d^*)^2}{\|x(0)\|^2}\geq\epsilon^2\Big)\nonumber\\
&&\leq\epsilon^{-2}\frac{\mathbb{E}\Big[\|x(k)-x^*\|^2\Big]-{d^*}^2}{\|x(0)\|^2} 
\leq \epsilon^{-2}\frac{C_8\sqrt{\gamma}^k+C_7-{d^*}^2}{\|x(0)\|^2}.
\end{eqnarray}
From \eqref{probability_inequality}, it follows that,
\begin{eqnarray}\label{shart_baraye_Tav}
&&\epsilon^{-2}x_{\min}^{-2}(C_8\sqrt{\gamma}^k+C_7-{d^*}^2)\leq\epsilon\Rightarrow\nonumber\\
&&\text{Pr}\Big(\frac{\|x(k)-x^*\|-d^*}{\|x(0)\|}\geq\epsilon\Big)\leq\epsilon.
\end{eqnarray}
Using  Definition~\ref{epsilon_averaging} and Assumption~\ref{non_zero_minimum_value}, one can obtain a lower bound for $N_\text{av}(\epsilon)$ from \eqref{shart_baraye_Tav},
\begin{equation}\label{T_av_asli}
N_\text{av}(\epsilon)\geq\frac{\log \frac{a}{\epsilon^{3}-b}}{\log\frac{1}{\sqrt{\gamma}}},
\end{equation}
where $a:=C_8x_{\min}^{-2}$ and $b:=(C_7-{d^*}^2)x_{\min}^{-2}$. By Lemma~\ref{gamma_less_1}, \eqref{d*<C7} and the condition on $\phi$, $0<\phi<1$, $a$ and $b$ are positive and increasing functions of $\gamma$.
$\hfill\blacksquare$

\begin{remark}The lower bound for $N_\text{av}(\epsilon)$ is an increasing function of $\gamma$ defined in Lemma~\ref{gamma_less_1}. Therefore, as $\gamma$ increases,  more  iterations  are required to converge to a Nash equilibrium. 
$d^*$ is not dependent on the topology of the interference and communication graphs as long as Assumptions~1 and 2 are met. Thus, $d^*$ does not change with $\gamma$. \end{remark}
\begin{remark}
As  \eqref{T_av_asli} shows, since $\epsilon^3-b$ has to be positive we are not able to find a lower bound for the number of iterations for any value of $\epsilon$. However, for the case when ${d^*}^2=C_7$, $N_\text{av}(\epsilon)$ is the minimum number of iterations it takes for $\|x(k)-x^*\|$ to approach an $\epsilon$-ball around $\sqrt{C_7}$, and $N_\text{av}(\epsilon)\geq\frac{\log \frac{a}{\epsilon^{3}}}{\log\frac{1}{\sqrt{\gamma}}}$. Thus as  $\gamma$ increases, $N_\text{av}(\epsilon)$ increases for any value of $\epsilon$.
\end{remark}
Characterizing $\gamma$ and its relation with the communication and interference graphs shed light on the  convergence rate of the algorithm. The following lemma characterizes $\gamma$.
\begin{lemma}\label{gamma=lambda2}
Let $\bar{W}:=\mathbb{E}[W(k)]$ be the expected communication matrix. Then $\bar{W}$ is doubly stochastic with $\lambda_{\max}(\bar{W})=1$. Let $\lambda_2(\bar{W})$ be the second largest eigenvalue of $\bar{W}$, i.e., $\lambda_2(\bar{W}):=\max_{\lambda\neq 1}\lambda(\bar{W})$. Then $\gamma$ as defined in Lemma~\ref{gamma_less_1}, $\gamma=\lambda_{\max}\big(\mathbb{E}[Q(k)^TQ(k)]\big)$,  satisfies $\gamma=\lambda_2(\bar{W})$.
\end{lemma}
\par{\emph{Proof}}. See Appendix.\\
By Lemma~\ref{gamma=lambda2}, $\gamma$ is the second largest eigenvalue of $\bar{W}$, hence  $N_\text{av}(\epsilon)$ depends on the structure of the expected communication matrix. It can be seen in the proof of Lemma~\ref{gamma=lambda2} that $\bar{W}$ depends on parameters associated with the interference and the communication graphs. In general it is difficult to see the effect of each parameter on $\lambda_2$ of $\bar{W}$. Note that for an irreducible doubly stochastic $\bar{W}$, we can find an upper bound for the spectral gap (i.e. $1-\lambda_2$), \cite{fiedler1972bounds}. However, $\bar{W}$ could be reducible which prevents us from using such a powerful tool. Later on in the simulation section, we will compute $\gamma$ numerically and we will see the effect of $\gamma$ on the convergence rate. Another parameter that $N_\text{av}(\epsilon)$ \eqref{shart_baraye_Tav} is dependent on is $\phi$, \eqref{condition}, which is associated with the cost functions.

To sum up, from the perspective of parameters associated with $G_I$, we can conclude that each iteration is shorter when  the interference graph is considered. Moreover, the number of iterations is tightly dependent on the second largest eigenvalue of the expected communication matrix hence on $G_C$.
\color{black}\section{Simulation Results}
In this section we present a numerical example and compare Algorithm~1 and Algorithm~2. Consider a \emph{Wireless Ad-Hoc Network} (WANET) which consists of 16 mobile nodes interconnected by multi-hop communication paths \cite{jayash7}. Consider $N_{ah}=\{1,\ldots,16\}$ as the set of wireless nodes and $L_{ah}=\{L_l\}_{l\in\mathcal{L}}$ as the set of links connecting the nodes,  $\mathcal{L}=\{1,\ldots,16\}$  the set of link indices. Let $V=\{U_1,\ldots,U_{15}\}$ denote the set of users (players) who want to use this wireless network to transfer data. Fig.~3~(a) represents the topology of the WANET in which a unique path is assigned to each user to transfer his data from the source to the destination node. Each $U_i$ is characterized by a set of links (path), $R_i$, $i\in V$.
\begin{figure}
\vspace{-1cm}
\hspace{-2.7cm}
\centering
\includegraphics [scale=0.5]{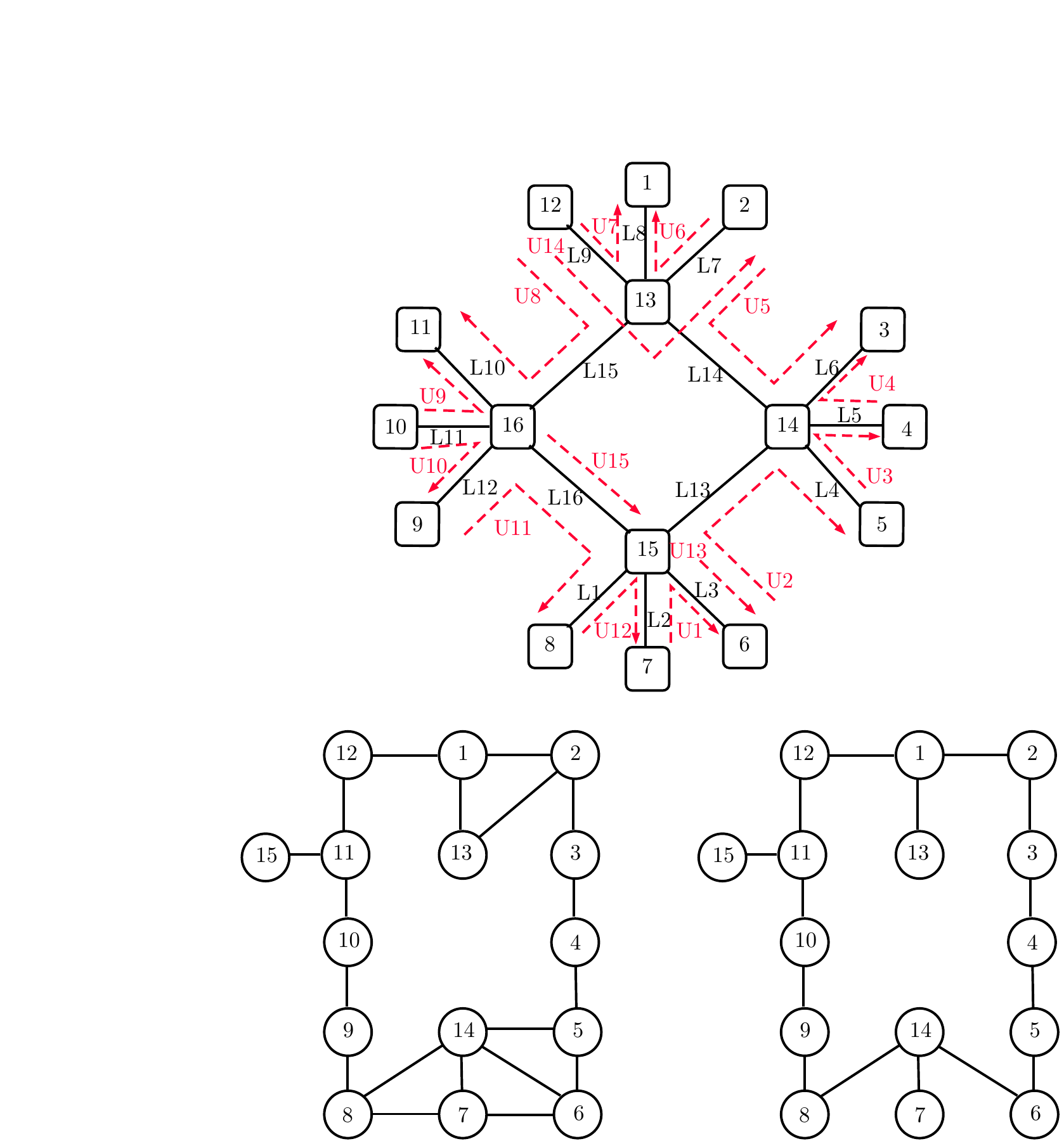}
\caption{(a)  Wireless Ad-Hoc Network. (b)  Interference graph  $G_I$ (the bottom left figure). (c) Communication graph $G_C$ (the bottom right figure).}
\end{figure}
The interferences between users  are represented in Fig.~3~(b). Nodes specify the users and edges show which users have a common link in their paths.
Each link $L_j\in L_{ah}$ has a positive capacity $C_j>0$, $j\in\mathcal{L}$. Each $U_i$, $U_i\in V$, sends a non-negative flow $x_i$, $0\leq x_i\leq 10$, over $R_i$, and has a cost function $J_i$  defined as
\begin{equation*}\label{Cost_fcn_gen}
J_i(x_i,x_{-i}^i):=\sum_{j:L_j\in R_i}\frac{\kappa}{C_j-\sum_{w:L_j\in R_w}x_w}-\chi_i \log(x_i+1),\end{equation*}
where $\kappa$ is a positive network-wide known parameter and $\chi_i$ is a positive user-specific parameter. The notation $a:b\in c$ translates into \enquote{set of $a$'s such that $b$ is contained in $c$}.
We run Algorithm~1 over the communication graph $G_C$ in Fig.~3~(c), and compare its convergence rate with that of  Algorithm~2 over the same $G_C$.
Let $\chi_i=10$, $i\in V$ and $C_j=10$,  $j\in\mathcal{L}$.
\begin{figure}
\vspace{-2.15cm}
\hspace{-1.9cm}
\centering
\begin{minipage}[b]{0.45\linewidth}
\includegraphics [scale=0.265]{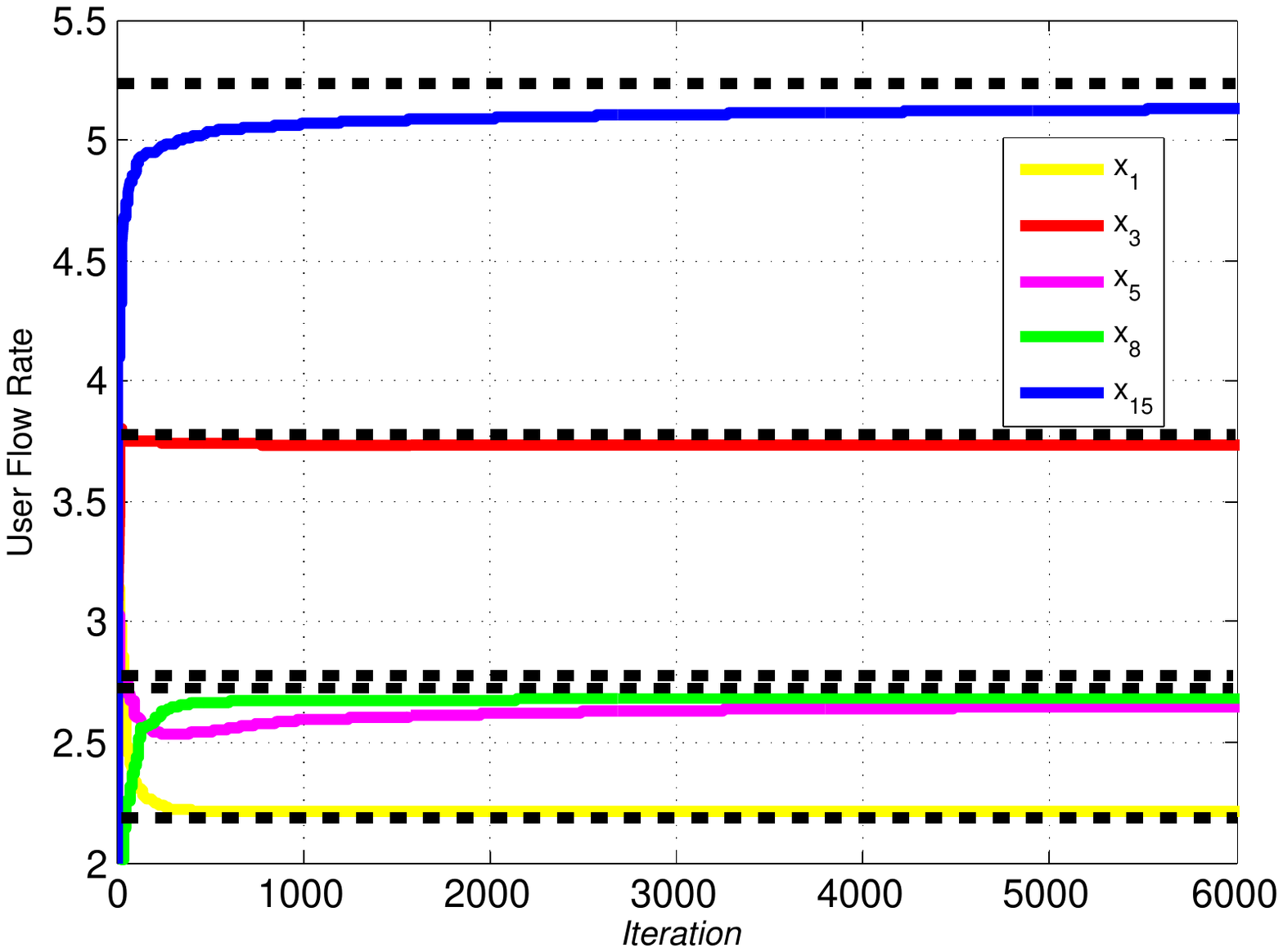}
\label{fig:minipage1}
\end{minipage}
\hspace{0.15cm}
\begin{minipage}[b]{0.45\linewidth}
\includegraphics [scale=0.265]{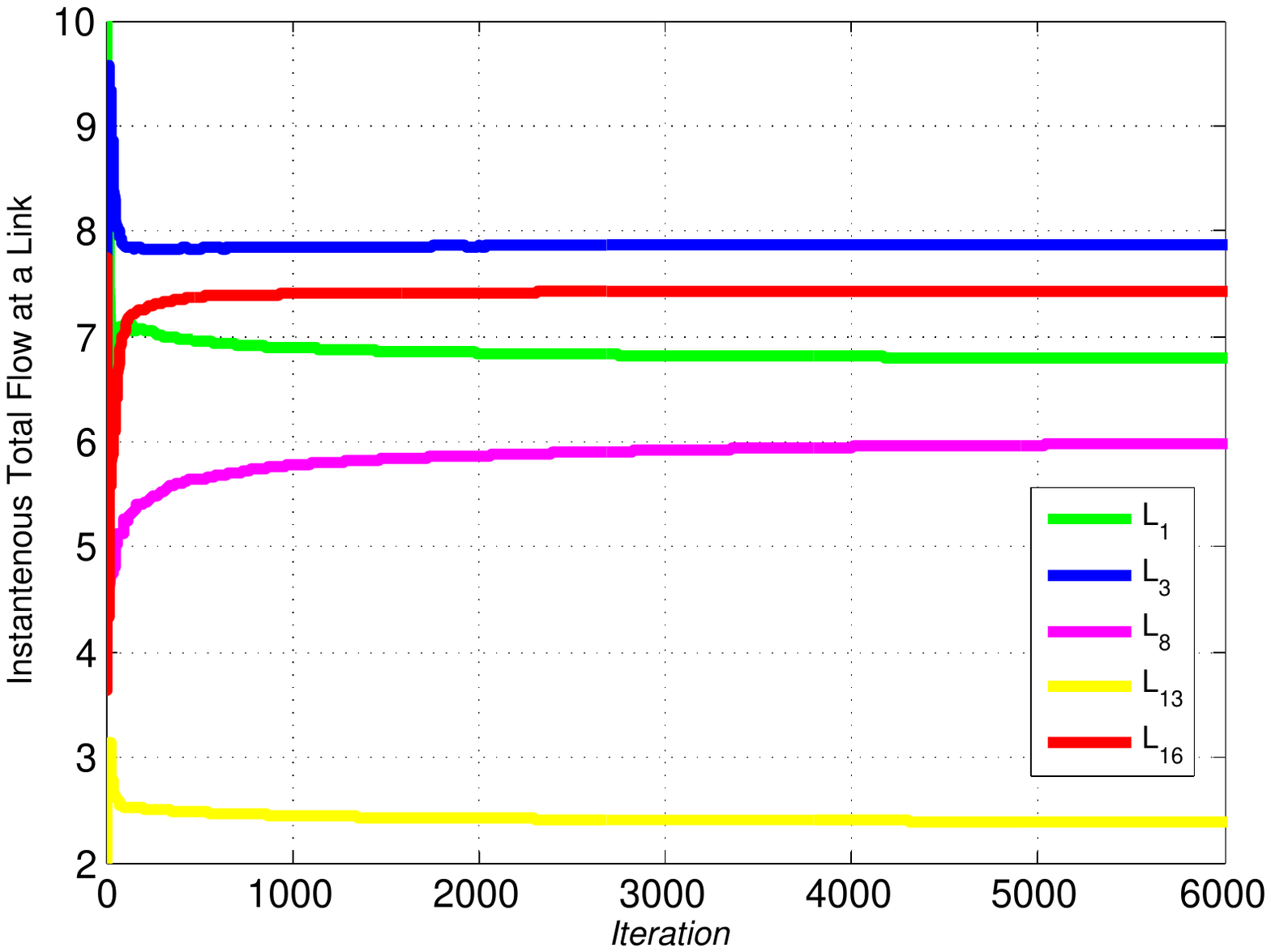}
\label{fig:minipage2}
\end{minipage}
\vspace{-2.7cm}
\caption{Flow rates of the selected users and total flow rates at the selected links by Algorithm~1.}
\end{figure}
\begin{figure}
\vspace{-2.35cm}
\hspace{-1.9cm}
\centering
\begin{minipage}[b]{0.45\linewidth}
\includegraphics [scale=0.265]{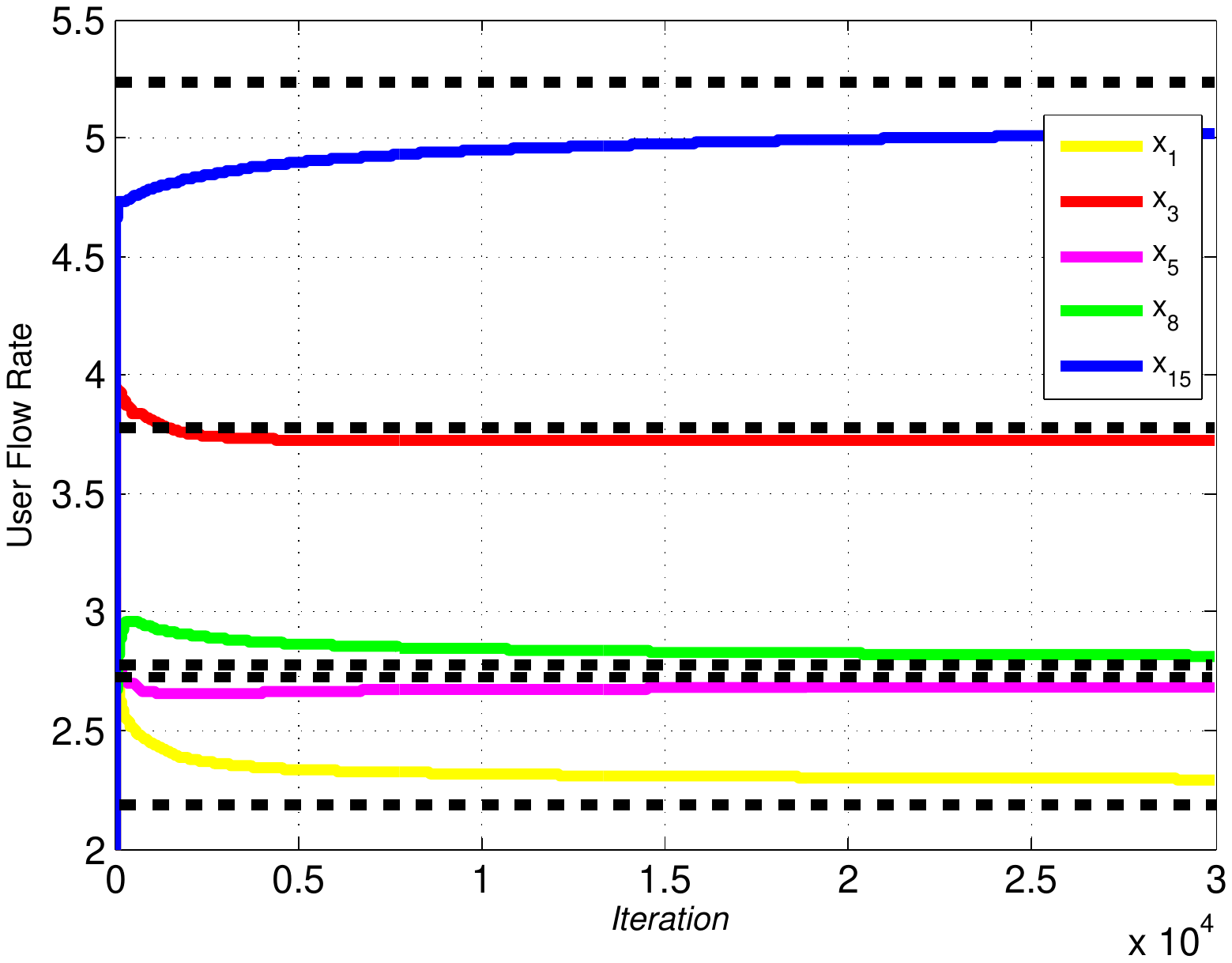}
\label{fig:minipage1}
\end{minipage}
\hspace{0.15cm}
\begin{minipage}[b]{0.45\linewidth}
\includegraphics [scale=0.265]{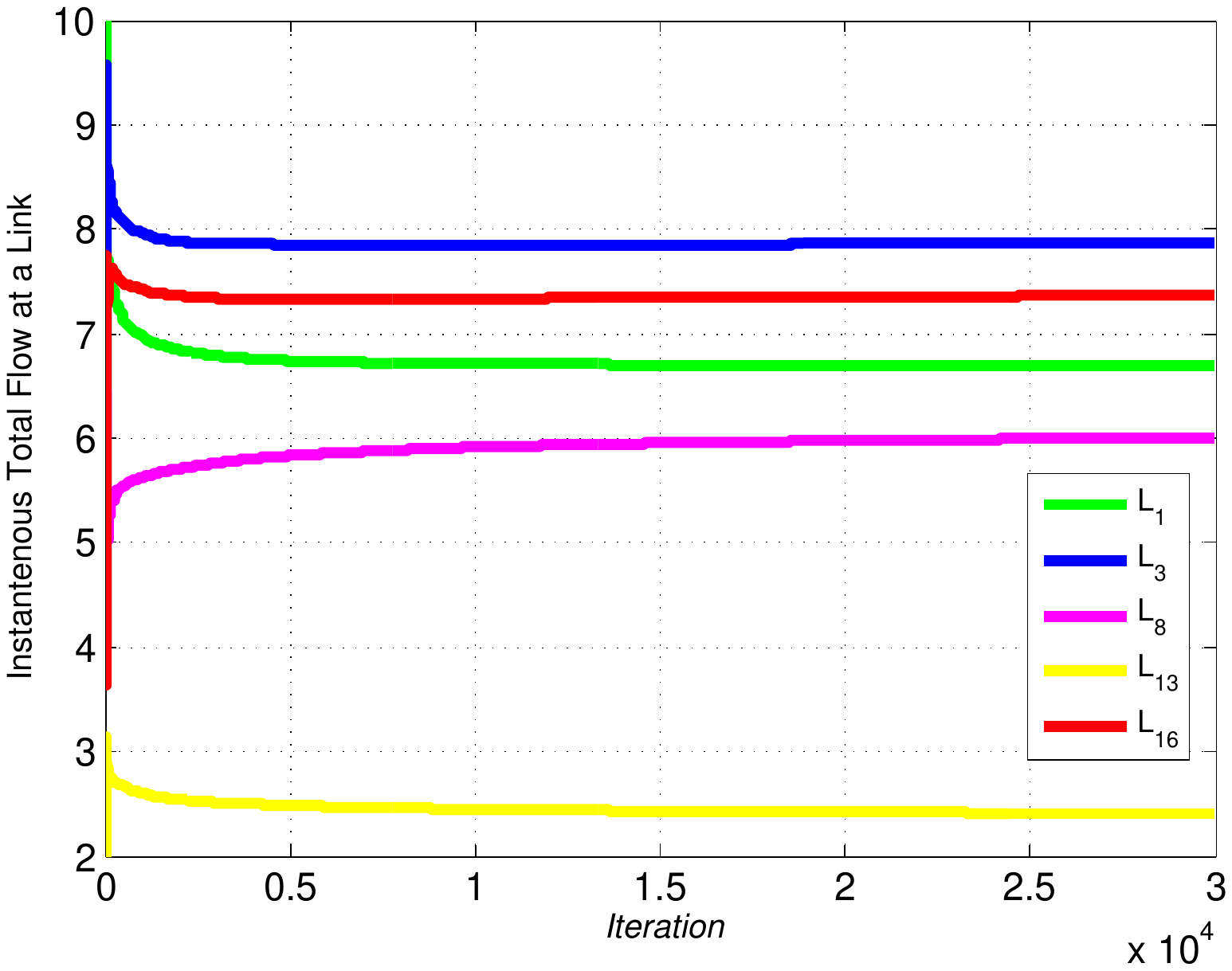}
\label{fig:minipage2}
\end{minipage}
\vspace{-2.7cm}
\caption{Flow rates of the selected users and total flow rates at the selected links by Algorithm~2.}
\vspace{-0.25cm}
\end{figure}
Fig.~4 and Fig~5 show convergence of Algorithm~1 and Algorithm~2 for diminishing step sizes. The dashed lines represent the Nash equilibrium.  The normalized error ($\frac{\|x-x^*\|}{\|x^*\|}\times100\%$) is 3.93\% after 6000 iterations for Algorithm~1, and after 30000 iterations for  Algorithm~2. Algorithm~1 needs 5 times fewer iterations, and each iteration is 6 times shorter. Thus, Algorithm~1 is 30 times faster than Algorithm~2 in this example.  In order to verify the analysis in Section~VI (Theorem~\ref{theorem_convergence_rate}), we run Algorithms~1 and 2 with constant step sizes $\alpha_{k,i}=0.1$. We compute $\gamma$ for Algorithms~1,2 using Lemma~\ref{gamma=lambda2}:  $\gamma_1=0.983$, $\gamma_2=0.994$. Fig.~6 shows how the lower bound on $N_\text{av}(\epsilon)$,  \eqref{N_av},  varies with  $\gamma$,  for $\epsilon=0.01$.
\begin{figure}
\vspace{-2.5cm}
\centering
\includegraphics [scale=0.33]{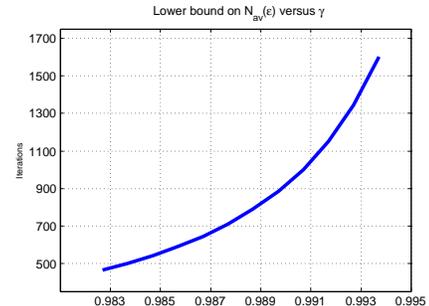}
\label{fig11111}
\vspace{-2.8cm}
\caption{Lower bound on $N_\text{av}(\epsilon)$ for $\alpha_{k,i}=0.1$ and $\epsilon=0.01$ versus $\gamma$.}
\vspace{-0.2cm}
\end{figure}
This confirms that Algorithm~2 (with greater $\gamma$) requires more iterations to converge to a Nash equilibrium.
\section{Conclusions}
In this paper we proposed a gossip algorithm to find a Nash equilibrium in a networked game. An interference graph is used to illustrate the locality of the cost functions. Players exchange only their required information over a communication graph. We proved convergence to a Nash equilibrium. 
We also presented convergence proof for the case when the component functions are not known by the players. Moreover, we showed the effect of the second largest eigenvalue of the expected communication matrix on the convergence rate. 
\color{black}\section*{Appendix}
\vspace{-0.2cm}{\color{black}\par{\emph{Proof of Lemma~\ref{tria_lemma}}}. In the following, we prove $N_I(i)\subseteq\bigcup_{j\in N_C(i)}\tilde{N}_I(j)$ for $i\in V$ from which it is straightforward to deduce \eqref{Lemma_set}.

For the case when $G_C=G_I$, we obtain,
\begin{equation}\label{Lemma_G_C=G_I}
\bigcup_{j\in N_C(i)}\!\tilde{N}_I(j)\!=\!\bigcup_{j\in N_I(i)}\!\tilde{N}_I(j)\!\supseteq\!\bigcup_{j\in {N}_I(i)}\!\{j\}\!=\!N_I(i).
\end{equation}
In \eqref{Lemma_G_C=G_I}, we used the fact that $\{j\}\subseteq\tilde{N}_I(j)$ by the definition of $\tilde{N}_I(j)$.

Now assume that $G_m\subseteq G_C\subset G_I$. To prove \eqref{Lemma_set}, it is sufficient to show that $N_I(i)\subseteq\bigcup_{j\in N_m(i)}\tilde{N}_m(j)$, where $N_m(i)$ is the set of neighbors of player $i$ in $G_m$ and $\tilde{N}_m(i)$ in addition to $N_m(i)$ contains $\{i\}$. In other words we need to show that any neighbor of player $i$ in $G_I$ (any vertex with a path of length 1 away from $i$ in $G_I$) is either a neighbor or a \enquote{neighbor of a neighbor} of player $i$ in $G_m$ (a vertex with a path of length 1 or 2 away from $i$ in $G_m$). 

We prove this claim by contradiction: Assume that there exists $j\in N_I(i)$ such that $j\notin N_m(i)$ and $j\notin N_m(l)$ $\forall l\in N_m(i)$. Then, there exists no path of length 1 or 2 between $i$ and $j$ in $G_m$. Thus, adding that missing edge to $G_m$ does not form a triangle which violates the maximal triangle free property of $G_m$. This is a contradiction and hence our claim is true.
$\hfill\blacksquare$}
\par{\emph{Proof of Lemma~\ref{lemma_ext_stoch} Part $i)$}}. Using the definition of $W(k)$ \eqref{W_def} we obtain,
\begin{eqnarray}\label{W^T(k)W(k)}
\hspace{-0.2cm}&&W^T(k)W(k)
=I_m-\sum_{l\in\text{ind}(i_k,j_k)}(E_l^{i_k}-E_l^{j_k})(E_l^{i_k}-E_l^{j_k})^T\nonumber\\
\hspace{-0.2cm}&&+\frac{1}{4}\Big(\sum_{l\in\text{ind}(i_k,j_k)}(E_l^{i_k}-E_l^{j_k})(E_l^{i_k}-E_l^{j_k})^T\Big)^2=W(k),\nonumber
\end{eqnarray}
where we used $(E_{l}^{i_k}-E_{l}^{j_k})^T(E_{l^{\prime}}^{i_k}-E_{l^{\prime}}^{j_k})=\begin{cases} 2,&\text{if }l=l^{\prime}\\0,& \text{if }l\neq l^{\prime}\end{cases}$.\vspace{-0.4cm}
\par{\emph{Proof of Part $ii)$}}. Using the definitions of $H$ and $W(k)$ \eqref{H}, \eqref{W_def}, we expand $W(k)H$ as
\begin{eqnarray}
  \hspace{-0.2cm}&&W(k)H=\bigg[\sum_{i=1}^{N}E_j^i-\frac{1}{2}\sum_{l\in\text{ind}(i_k,j_k)}\bigg((E_l^{i_k}-E_l^{j_k})\nonumber\\
\hspace{-0.2cm}&&.\sum_{i=1}^{N}(E_l^{i_k}-E_l^{j_k})^TE_j^i\bigg)\bigg]_{j\in V}.\nonumber
\end{eqnarray}
Note that ${E_l^{i_k}}^TE_j^i=1$ if $i=i_k$, $j=l$ and ${E_l^{i_k}}^TE_j^i=0$ otherwise. Similarly ${E_l^{j_k}}^TE_j^i=1$ if $i=j_k$, $j=l$ and ${E_l^{j_k}}^TE_j^i=0$ otherwise. Consequently, $\sum_{i=1}^{N}({E_l^{i_k}}^T-{E_l^{j_k}}^T)E_j^i=0$ for all $j\in V$ and \eqref{W(k)H=Hlemma} is immediately obtained. Part $iii$ follows similarly.
$\hfill\blacksquare$
\par{\emph{Proof of Lemma~\ref{QZ=0}}}. From \eqref{W(k)H=Hlemma} in Lemma~\ref{lemma_ext_stoch}, \eqref{ave_Z} and the definition of $\bar{H}$ \eqref{Hbar}, we obtain,\vspace{-0.4cm}
\begin{eqnarray}
&&Q(k)Z(k)= (W(k)-H\bar{H}W(k))H\bar{H}\tilde{x}(k)\nonumber\\
        &&= (W(k)H\bar{H}-H\bar{H}W(k)H\bar{H})\tilde{x}(k)\nonumber\\
        &&= (H\bar{H}-H\text{diag}(1./\textbf{m})H^TW(k)H\text{diag}(1./\textbf{m})H^T)\tilde{x}(k).\nonumber
\end{eqnarray}
By Lemma~\ref{lemma_ext_stoch} Part $ii$ and \eqref{H}, one can verify that,
\begin{equation}\label{H^TH=diag(m)}
H^TW(k)H=H^TH=\text{diag}(\textbf{m}).
\end{equation}
This yields,
$Q(k)Z(k)= 
(H\bar{H}-H\text{diag}(1./\textbf{m})H^T)\tilde{x}(k)
=(H\bar{H}-H\bar{H})\tilde{x}(k)=\textbf{0}_m$.
$\hfill\blacksquare$
\vspace{-0cm}\par{\emph{Proof of Lemma~\ref{R=1}}}. By the definition of $\bar{H}$ \eqref{Hbar} and $H^TH=\text{diag}(\textbf{m})$ \eqref{H^TH=diag(m)}, one can obtain,
\begin{equation}
R^TR = (I_m-H\bar{H})^T(I_m-H\bar{H})=I_m-H\text{diag}(1./\textbf{m})H^T.\nonumber
\end{equation}
To find the eigenvalues of $R^TR$, we solve the following characteristic equation for $\lambda$.
\begin{eqnarray}\label{charac_eq}
  &&\text{det}(R^TR-\lambda I_m) = 0\nonumber\\
  &&\Leftrightarrow\text{det}((1-\lambda) I_m-H\text{diag}(1./\textbf{m})H^T) = 0.
\end{eqnarray}
We claim that  there exists at least one eigenvalue equal to 1, and all the eigenvalues of $R^TR$ are either 1 or 0. We prove the first claim in the following.
By \eqref{H}, we have the following for $\lambda=1$:
\begin{eqnarray}
&&\text{det}((1-\lambda) I_m-H\text{diag}(1./\textbf{m})H^T) \nonumber\\
&&= (-1)^m\text{det}\bigg(\sum_{j=1}^{N}\big(\frac{1}{m_j}\sum_{i=1}^{N}E_j^i\sum_{i=1}^{N}{E_j^i}^T\big)\bigg).\nonumber
\end{eqnarray}
By critically observing $\sum_{j=1}^{N}\big(\frac{1}{m_j}\sum_{i=1}^{N}E_j^i\sum_{i=1}^{N}{E_j^i}^T\big)$, it follows that this matrix consists of the rows $\frac{1}{m_j}\sum_{i=1}^{N}{E_j^i}^T$ each repeated $m_j$ times for $j\in V$. By Remark~\ref{mi>1} ($m_j>1$, $\forall j\in V$), the repetition of each row is at least 2 times, thus $\text{det}\big(\sum_{j=1}^{N}\big(\frac{1}{m_j}\sum_{i=1}^{N}E_j^i\sum_{i=1}^{N}{E_j^i}^T\big)\big)=0$ and there exists at least one $\lambda=1$.
Next, we prove the claim that all  non-one eigenvalues are 0. For now, assume that there exist $k$ eigenvalues, $k\in\mathbb{N}$, $1\leq k\leq m$, equal to 1 for $R^TR$. For $\lambda\neq 1$, we have,
  \begin{eqnarray}\label{lambda_m-k}
  &&\text{det}((1-\lambda) I_m-H\text{diag}(1./\textbf{m})H^T) = 0\Leftrightarrow\nonumber\\
  &&(1-\lambda)^{m-k}\text{det}(I_m-\frac{1}{1-\lambda}H\text{diag}(1./\textbf{m})H^T) = 0.\nonumber
\end{eqnarray}
Multiplying and dividing by $\text{det}(\text{diag}(\textbf{m}))=\prod_{i=1}^{N}m_i$ (non-zero  by Remark~\ref{mi>1}), yields,
\begin{eqnarray}\label{lambda_m-k}
&&\frac{(1-\lambda)^{m-k}}{\text{det}(\text{diag}(\textbf{m}))}\, \text{det}(\text{diag}(\textbf{m}))\text{det}(I_m-\frac{1}{1-\lambda}\nonumber\\
&&.H\text{diag}(1./\textbf{m})H^T) = 0.
\end{eqnarray}
Let $A$, $B$ and $X$ be $m\times n$, $n\times m$ and $m\times m$ matrices, respectively. Let also $X$ be an invertible matrix. By the \emph{generalized Sylvester's determinant} theorem, the following equality holds:
\begin{eqnarray}\label{sylvester}
&&\text{det}(X+AB)=\text{det}(X)\, \text{det}(I_n+BX^{-1}A).
\end{eqnarray}
Let $X:=\text{diag}(\textbf{m})$, $A:=H^T$ and $B:=-\frac{H}{1-\lambda}$, so that  from \eqref{sylvester} and \eqref{lambda_m-k} we obtain,
\begin{eqnarray}
&&\frac{(1-\lambda)^{m-k}}{\prod_{i=1}^{N}m_i}\text{det}(\text{diag}(\textbf{m})-\frac{H^TH}{1-\lambda}) = 0.\nonumber
\end{eqnarray}
Since $H^TH=\text{diag}(\textbf{m})$ \eqref{H^TH=diag(m)}, this is equivalent to
\begin{eqnarray}\label{lamda=0}
\hspace{-0.2cm}&&\frac{(1-\lambda)^{m-k}}{\prod_{i=1}^{N}m_i}\text{det}(\frac{-\lambda}{1-\lambda}\text{diag}(\textbf{m})) = 0 \, \Leftrightarrow
(-\lambda)^{m-k} = 0.\nonumber
\end{eqnarray}
This verifies that there exist $m-k$ eigenvalues, $0\leq m-k<m$, equal to 0 as we claimed. Thus, $\lambda_{\text{max}}(R^TR)=1$ and $\|R\|=1$.
$\hfill\blacksquare$
\par{\emph{Proof of Lemma~\ref{gamma_less_1}}}. Using Lemma~\ref{lemma_ext_stoch} and \eqref{H^TH=diag(m)}, we obtain,\vspace{-0.3cm}
\begin{equation}
  Q(k)^TQ(k)=W(k)-H\bar{H}.\nonumber
\end{equation}
Since $\mathbb{E}[Q(k)^TQ(k)]$ is an $m\times m$ symmetric matrix, then,\vspace{-0.3cm}
\begin{equation}
\gamma=\sup_{x\in\mathbb{R}^m,\|x\|=1}x^T\mathbb{E}[Q(k)^TQ(k)]x.\nonumber
\end{equation}
We find $\gamma$ for the vector subspace $\{x\in\mathbb{R}^m:x=c\textbf{1}_m,c\in\mathbb{R}\}$ and moreover, for its orthogonal complement vector subspace $\{x\in\mathbb{R}^m:x^T\textbf{1}_m=0\}$.
Let for now $x=c\textbf{1}_m$. By the doubly stochastic property of $W(k)$ we obtain,
\begin{eqnarray}
\gamma&=&c^2\mathbb{E}[\textbf{1}_m^TQ(k)^TQ(k)\textbf{1}_m]\nonumber\\
&=& c^2\mathbb{E}[\textbf{1}_m^TW(k)^TW(k)\textbf{1}_m]-c^2\mathbb{E}[\textbf{1}_m^TH\bar{H}\textbf{1}_m]\nonumber\\
&=&c^2\textbf{1}_m^T\textbf{1}_m-c^2\textbf{1}_m^TH\text{diag}(1./\textbf{m})H^T\textbf{1}_m.\nonumber
\end{eqnarray}
Note that from the definition of $H$ \eqref{H} we obtain 
$\textbf{1}_m^TH=\textbf{m}^T$.
Then,
\begin{equation}\label{c1_eig}
\gamma=c^2m-c^2\textbf{m}^T\text{diag}(1./\textbf{m})\textbf{m}=c^2m-c^2m=0.
\end{equation}
Thus, $\gamma=0$ for $x=c\textbf{1}_m$. Consider now that $x$ belongs to $\{x\in\mathbb{R}^m:x^T\textbf{1}_m=0\}$ and compute
\begin{eqnarray}\label{gamma=sup}
\hspace{-0.2cm}&&\gamma=\sup_{\substack{x\in\mathbb{R}^m,\\\|x\|=1,\\x^T\textbf{1}_m=0}}x^T\mathbb{E}[Q(k)^TQ(k)]x=\sup_{\substack{x\in\mathbb{R}^m,\\\|x\|=1,\\x^T\textbf{1}_m=0}}\mathbb{E}[x^TQ(k)^TQ(k)x]\nonumber\\
\hspace{-0.2cm}&&=\sup_{\substack{x\in\mathbb{R}^m,\\\|x\|=1,\\x^T\textbf{1}_m=0}}\big(\mathbb{E}[\|W(k)x\|^2]-x^TH\bar{H}x\big).
\end{eqnarray}
Note that $\|W(k)x\|^2\leq\|x\|^2=1$. As a result
\begin{equation}\label{E[W(k)x^2] 1}
\mathbb{E}[\|W(k)x\|^2]\leq 1.
\end{equation}
Note also that the minimum eigenvalue of $H\bar{H}$ is 0 and the associated normalized eigenvector  $v_\text{min}$ satisfies $H\bar{H}v_\text{min}=0$, $v_\text{min}^T\textbf{1}_m=0$ and $\|v_\text{min}\|=1$. Then we achieve,
\begin{equation}\label{vminhhvmin}
\inf_{\substack{x\in\mathbb{R}^m,\\\|x\|=1,\\x^T\textbf{1}_m=0}}x^TH\bar{H}x=v_\text{min}^TH\bar{H}v_\text{min}=0.
\end{equation}
By \eqref{E[W(k)x^2] 1} and \eqref{vminhhvmin}, it is straightforward to show using \eqref{gamma=sup} that $\gamma\leq 1$. We now claim that $\gamma<1$. To prove our claim we need to verify
\begin{equation}\label{EWvneq1}
\mathbb{E}[\|W(k)v_\text{min}\|^2]\neq 1.\nonumber
\end{equation}
By \eqref{W_def} we obtain,
\begin{eqnarray}
&&\mathbb{E}[\|W(k)v_\text{min}\|^2]\nonumber\\
&&=\mathbb{E}\big[v_\text{min}^T\big(I_m-\frac{1}{2}\sum_{l\in\text{ind}(i_k,j_k)}(E_l^{i_k}-E_l^{j_k})(E_l^{i_k}-E_l^{j_k})^T\big)v_\text{min}\big]\nonumber\\
&&=1-\frac{1}{2}\mathbb{E}\big[\sum_{l\in\text{ind}(i_k,j_k)}\big\|(E_l^{i_k}-E_l^{j_k})^Tv_\text{min}\big\|^2\big].\nonumber
\end{eqnarray}
We expand the expected value considering that each player $i\in V$ communicates with player $j\in N_C(i)$ with probability $\frac{1}{\text{deg}_{G_C}(i)}$.
\begin{eqnarray}\label{EWvmin}
&&\mathbb{E}[\|W(k)v_\text{min}\|^2]=1-\frac{1}{2N}\sum_{i\in V}\frac{1}{\text{deg}_{G_C}(i)}\nonumber\\
&&.\sum_{j\in N_C(i)}\sum_{l\in\text{ind}(i,j)}\big\|(E_l^{i}-E_l^{j})^Tv_\text{min}\big\|^2.
\end{eqnarray}
By contradiction, we assume that $\mathbb{E}[\|W(k)v_\text{min}\|^2]=1$. From \eqref{EWvmin}, it follows that $(E_l^{i}-E_l^{j})^Tv_\text{min}=0$, for all $i\in V$, $j\in N_C(i)$ and $l\in\text{ind}(i,j)$. {\color{black}Moreover, by Lemma~\ref{tria_lemma}, we simply have $\bigcup_{j\in N_C(i)}\text{ind}(i,j)=\tilde{N}_I(i)$ which shows that the indices $l\in\text{ind}(i,j)$ for every $i\in V$, $j\in N_C(i)$ span all the elements of $v_{\min}$.} Thus, for $\forall i\in V$, $\forall j\in N_C(i)$ and $l\in\text{ind}(i,j)$, we obtain,
\begin{equation}\label{contrad1}
(E_l^{i}-E_l^{j})^Tv_\text{min}=0\Rightarrow v_{\text{min}_z}=v_{\text{min}_r},
\end{equation}
where $z,r\in\{t:t=s_{in}, i\in V, n\in \tilde{N}_I(i)\},\,z\neq r$ where $s_{in}$ is as defined in \eqref{s_ij}. On the other hand, by \eqref{vminhhvmin}, we obtain,
\vspace{-0.3cm}
\begin{eqnarray}\label{contrad2}
&&v_\text{min}^TH\bar{H}v_\text{min}=0\Leftrightarrow v_\text{min}^T\big[\sum_{i=1}^{N}E_1^i,\ldots,\sum_{i=1}^{N}E_N^i\big]\text{diag}(1./\textbf{m})\nonumber\\
&&.\big[\sum_{i=1}^{N}{E_1^i}^T,\ldots,\sum_{i=1}^{N}{E_N^i}^T\big]^Tv_\text{min}=0\Leftrightarrow\nonumber\\
&&\sum_{j=1}^{N}\frac{1}{m_j}\big\|\sum_{i=1}^{N}{E_j^i}^Tv_\text{min}\big\|^2=0\Leftrightarrow\sum_{i=1}^{N}{E_j^i}^Tv_\text{min}=0,\forall j\in V\nonumber\\
&&\Leftrightarrow\sum_{\substack{z\in\{t:t=s_{in}, i\in V, n\in \tilde{N}_I(i)\}}}v_{\text{min}_z}=0.
\end{eqnarray}
From \eqref{contrad1} and \eqref{contrad2} we obtain that $v_{\text{min}_z}=0$ for $z\in\{1,\ldots,m\}$ which contradicts $\|v_\text{min}\|=1$. Thus, the assumption $\mathbb{E}[\|W(k)v_\text{min}\|^2]=1$ is false which implies $\mathbb{E}[\|W(k)v_\text{min}\|^2]<1$.
This concludes that $\gamma<1$ for the vector subspace $\{x\in\mathbb{R}^m:x^T\textbf{1}_m=0\}$. Moreover, \eqref{c1_eig} verifies that $\gamma=0$ for the orthogonal complement vector subspace $\{x\in\mathbb{R}^m:x=c\textbf{1}_m,c\in\mathbb{R}\}$ which completes the proof.
$\hfill\blacksquare$\vspace{-0.0cm}
{\color{black}
	\par{\emph{Proof of Theorem~\ref{consensus1} Part $i)$}}.\\
	\emph{\textbf{Procedure:} First, we find an upper bound for $\mathbb{E}\Big[\|\tilde{x}(k+1)-Z(k+1)\|\Big |\mathcal{M}_k\Big]$ and simplify it to a similar format as in Lemma~\ref{sigmond}. Then we verify the conditions of Lemma~\ref{sigmond} step by step and apply it to the achieved upper bound.}}

We derive an upper bound for $\mathbb{E}\Big[\|\tilde{x}(k+1)-Z(k+1)\|\Big |\mathcal{M}_k\Big]$. Using \eqref{temp_update}, \eqref{s_ij} and \eqref{x_hat_tor}, we obtain $\tilde{x}_j^i(k+1)=\hat{x}_j^i(k)=[\bar{x}_r(k)]_{r=s_{ij}}$ for $i\in V$, $j\in N_I(i)$, $i\neq j$ and $\tilde{x}_i^i(k+1)=x_i(k+1)$ for $i\in V$. Then we have,
\begin{equation}
\tilde{x}(k+1)=W(k)\tilde{x}(k)+\mu(k+1),
\end{equation}
where $\mu(k+1):=[(x_i(k+1)-\bar{x}_{s_{ii}}(k))e_i]_{i\in V}$. Using \eqref{ave_Z}, we obtain,
\begin{eqnarray}
&&\tilde{x}(k+1)-Z(k+1)\nonumber\\
&&=W(k)\tilde{x}(k)+\mu(k+1)-H\bar{H}\tilde{x}(k+1)\nonumber\\
&&=W(k)\tilde{x}(k)+\mu(k+1)-H\bar{H}W(k)\tilde{x}(k)-H\bar{H}\mu(k+1)\nonumber\\
&&=Q(k)\tilde{x}(k)+R\mu(k+1),
\end{eqnarray}
with $Q(k)$ and $R$ as in Lemma~\ref{QZ=0} and Lemma~\ref{R=1}, respectively. Using Lemma~\ref{QZ=0}, we arrive at
\begin{eqnarray}\label{term1&term2}
&&\tilde{x}(k+1)-Z(k+1)=Q(k)(\tilde{x}(k)-Z(k))+R\mu(k+1)\nonumber\\
&&\Rightarrow\mathbb{E}\Big[\|\tilde{x}(k+1)-Z(k+1)\|\Big |\mathcal{M}_k\Big]=\\
&&\underbrace{\mathbb{E}\Big[\|Q(k)(\tilde{x}(k)-Z(k))\|\Big |\mathcal{M}_k\Big]}_{\text{Term~1}}+\underbrace{\mathbb{E}\Big[\|R\mu(k+1)\|\Big |\mathcal{M}_k\Big]}_{\text{Term~2}}.\nonumber
\end{eqnarray}
{\color{black}Let $\gamma=\lambda_{\max}\big(\mathbb{E}[Q(k)^TQ(k)]\big)$ be as in Lemma~\ref{gamma_less_1}. We obtain an upper bound for Term~1 in \eqref{term1&term2} as the following:
	\begin{eqnarray}\label{term1}
	\hspace{-0cm}\text{Term}~1&\leq&\sqrt{\mathbb{E}\Big[\|Q(k)(\tilde{x}(k)-Z(k))\|^2\Big |\mathcal{M}_k\Big]}\nonumber\\
	\hspace{-0cm}&\leq&\sqrt{\gamma}\|\tilde{x}(k)-Z(k)\|.
	\end{eqnarray}
	Note that by Lemma~\ref{gamma_less_1}, $\gamma<1$. This fact will be used as a key to bound Term~1.
	
	To bound Term~2, we use $\|R\|=1$ and $x_i(k+1)=x_i(k)=\bar{x}_{s_{ii}}$ for $i\notin\{i_k,j_k\}$. Then,
	\begin{eqnarray}\label{term2}
	&&\hspace{-0.1cm}\text{Term}~2\leq\mathbb{E}\Big[\|\mu(k+1)\|\Big |\mathcal{M}_{k}\Big]\nonumber\\
	&&\hspace{-0.1cm}=\mathbb{E}\Bigg[\sqrt{\sum_{i\in V}\|x_{i}(k+1)-\bar{x}_{s_{ii}}(k)\|^2}\Big |\mathcal{M}_{k}\Bigg]\nonumber\\
	&&\hspace{-0.1cm}=\mathbb{E}\Bigg[\sqrt{\sum_{i\in \{i_k,j_k\}}\|x_{i}(k+1)-\bar{x}_{s_{ii}}(k)\|^2}\Big |\mathcal{M}_{k}\Bigg]\nonumber\\
	&&\hspace{-0.1cm}\leq\mathbb{E}\Bigg[\sqrt{\sum_{i\in \{i_k,j_k\}}\Big(2\|x_{i}(k)-\bar{x}_{s_{ii}}(k)\|^2+2\alpha_{k,i}^2C^2}\Big)\Big |\mathcal{M}_{k}\Bigg]\nonumber\\
	&&\hspace{-0.1cm}\leq\sqrt{2}\mathbb{E}\Bigg[\sum_{i\in \{i_k,j_k\}}\|x_{i}(k)-\bar{x}_{s_{ii}}(k)\|\Big |\mathcal{M}_{k}\Bigg]+2\alpha_{k,\max}C
	\end{eqnarray}
	The second inequality is obtained by using \eqref{local_step}, the non-expansive property of projection, Assumption~\ref{assump} (equation \eqref{bounded}) and $(a+b)^2\leq2a^2+2b^2$. 
	
	We simplify \eqref{term2} using $\bar{x}_{s_{ii}}(k)=\frac{\tilde{x}_{i}^{i}(k)+\tilde{x}_{i}^{j}(k)}{2}$ for $i,j\in\{i_k,j_k\}$ which is deduced implicitely from \eqref{xbar}. Then,
	\begin{eqnarray}\label{term2}
	&&\hspace{-0.1cm}\text{Term}~2\leq\frac{\sqrt{2}}{2}\sum_{i\in \{i_k,j_k\}}\|\tilde{x}_{i}^i(k)-\tilde{x}_{i}^j(k)\|+2\alpha_{k,\max}C
	\end{eqnarray}
	From \eqref{term1&term2}, \eqref{term1} and \eqref{term2} it follows that 
	\begin{eqnarray}\label{term1&term2_karshode}
	&&\mathbb{E}\Big[\|\tilde{x}(k+1)-Z(k+1)\|\Big |\mathcal{M}_{k}\Big]\leq\sqrt{\gamma}\|\tilde{x}(k)-Z(k)\|\nonumber\\
	&&+\frac{\sqrt{2}}{2}\sum_{i\in\{i_k,j_k\}}\|\tilde{x}_{i}^i(k)-\tilde{x}_{i}^j(k)\|+2\alpha_{k,\max}C.
	\end{eqnarray}
	Multiplying the LHS and RHS of \eqref{term1&term2_karshode} by $\alpha_{k+1,\text{max}}$ and $\alpha_{k,\text{max}}$, respectively and spliting $\sqrt{\gamma}$ into $1$ and $1-\sqrt{\gamma}$, we obtain,
	\begin{eqnarray}\label{term1&term2_karshode_underbrace}
	&&\underbrace{\alpha_{k+1,\text{max}}\mathbb{E}\Big[\|\tilde{x}(k+1)-Z(k+1)\|\Big |\mathcal{M}_{k}\Big]}_{\mathbb{E}[V_{k+1}|\mathcal{M}_k]}\nonumber\\
	&&\leq\underbrace{\alpha_{k,\text{max}}\|\tilde{x}(k)-Z(k)\|}_{(1+u_k)V_k}-\underbrace{\alpha_{k,\text{max}}(1-\sqrt{\gamma})\|\tilde{x}(k)-Z(k)\|}_{\zeta_k}\nonumber\\
	&&+\underbrace{\frac{\sqrt{2}}{2}\alpha_{k,\text{max}}\sum_{i\in\{i_k,j_k\}}\|\tilde{x}_{i}^i(k)-\tilde{x}_{i}^j(k)\|+2\alpha_{k,\max}^2C}_{\beta_k}.
	\end{eqnarray} 
	We apply Lemma~\ref{sigmond} to \eqref{term1&term2_karshode_underbrace} to show that $\sum_{k=0}^{\infty}\alpha_{k,\text{max}}\|\tilde{x}(k)-Z(k)\|<\infty$ ($\sum_{k=0}^{\infty}\zeta_k<\infty$) a.s. We already assigned the parameters in \eqref{term1&term2_karshode_underbrace}. We need to verify that all the conditions of Lemma~\ref{sigmond} are met. Clearly $V_k\geq0$ and $u_k=0$.
	Using $\gamma<1$ (Lemma~\ref{gamma_less_1}) $\zeta_k>0$. To show $\sum_{k=0}^\infty\beta_k<\infty$ a.s., we use \eqref{diminish_step} and we need to verify that $\sum_{k=0}^{\infty}\alpha_{k,\max}\|\tilde{x}_{i}^{i}(k)-\tilde{x}_{i}^{j}(k)\|<\infty$ a.s. for $i,j\in\{i_k,j_k\}$.
	
	Using \eqref{local_step}, \eqref{temp_update}, \eqref{excluding}, projection's non-expansive property and \eqref{bounded}, it yields
	\begin{eqnarray}\label{ii-ij}
	\hspace{-0.0cm}\|\tilde{x}_{i}^{i}(k\!+\!1)\!-\!\tilde{x}_{i}^{j}(k\!+\!1)\|\!\leq\!\frac{1}{2}\|\tilde{x}_{i}^{i}(k)\!-\!\tilde{x}_{i}^{j}(k)\|\!+\!\alpha_{k,\max}C.
	\end{eqnarray}
	To obtain the inequality, we use $\tilde{x}_{i}^{j}(k+1)=\hat{x}_{i}^{j}(k)=\frac{\tilde{x}_{i}^{j}(k)+\tilde{x}_{i}^{i}(k)}{2}$ for $i,j\in\{i_k,j_k\}$. Take expected value of \eqref{ii-ij} and multiply its LHS and RHS by $\alpha_{k+1,\max}$ and $\alpha_{k,\max}$, respectively (since $\alpha_{k+1,\max}<\alpha_{k,\max}$).\vspace{-5pt}
	\begin{eqnarray}\label{alpha_trick1}
	&&\alpha_{k+1,\max}\mathbb{E}\Big[\|\tilde{x}_i^i(k+1)-\tilde{x}_i^j(k+1)\|\Big |\mathcal{M}_k\Big]\leq\\
	&&\alpha_{k,\max}\|\tilde{x}_i^i(k)-\tilde{x}_i^j(k)\|-\frac{\alpha_{k,\max}}{2}\|\tilde{x}_i^i(k)-\tilde{x}_i^j(k)\|+\alpha_{k,\max}^2C,\nonumber
	\end{eqnarray}
	where we split $\frac{\alpha_{k,\max}}{2}$ into $\alpha_{k,\max}$ and $-\frac{\alpha_{k,\max}}{2}$. Applying Lemma~\ref{sigmond} for  $V_{k}=\alpha_{k,\max}\|\tilde{x}_i^i(k)-\tilde{x}_i^j(k)\|$,$u_k=0$, $\beta_k=\alpha_{k,\max}^2C$,
	$\zeta_k=\frac{\alpha_{k,\max}}{2}\|\tilde{x}_i^i(k)-\tilde{x}_i^j(k)\|$, and using also  \eqref{diminish_step} and \eqref{bounded}  it follows that
	\begin{equation}
	\label{ii_ij<infty}
	\sum_{k=0}^{\infty}\alpha_{k,\max}\|\tilde{x}_{i}^{i}(k)-\tilde{x}_{i}^{j}(k)\|<\infty\quad\text{a.s.}
	\end{equation}
	
	Proof of Part $ii)$. The proof has similar steps as in the proof of Part $i$ and it is ommited due to space limitation.
	$\hfill\blacksquare$}
{\color{black}\par{\emph{Proof of Lemma~\ref{x_bar_converge}}}. By \eqref{xbar}, \eqref{ave_Z} and \eqref{W(k)H=Hlemma} we obtain,
\begin{eqnarray}
&&\sum_{k=0}^{\infty}\mathbb{E}\Big[\|\bar{x}(k)-Z(k)\|^2\Big |\mathcal{M}_k\Big]\nonumber\\
&&=\sum_{k=0}^{\infty}\mathbb{E}\Big[\|W(k)\tilde{x}(k)-Hz(k)\|^2\Big |\mathcal{M}_k\Big]\nonumber\\
&&=\sum_{k=0}^{\infty}\mathbb{E}\Big[\|W(k)\tilde{x}(k)-W(k)Hz(k)\|^2\Big |\mathcal{M}_k\Big]\nonumber\\
&&\leq\sum_{k=0}^{\infty}\mathbb{E}\Big[\|W(k)\|^2\|\tilde{x}(k)-Z(k)\|^2\Big |\mathcal{M}_k\Big],
\end{eqnarray}
where the second equality holds by \eqref{W(k)H=Hlemma}.
The proof follows from Theorem~\ref{consensus1} part $ii$ and the doubly stochastic property of $W(k)$ (Remark~\ref{barabari_W_ha}) which results $\|W(k)\|=1$.
$\hfill\blacksquare$}
\par{\emph{Proof of Lemma~\ref{gamma=lambda2}}}. Similar to the method used in \eqref{EWvmin}, we can derive the following for $\bar{W}$.
\begin{eqnarray}\label{Ewexpanded}
&&\bar{W}=\mathbb{E}[W(k)]=I_m-\frac{1}{2N}\sum_{i\in V}\frac{1}{\text{deg}_{G_C}(i)}\nonumber\\
&&.\sum_{j\in N_C(i)}\sum_{l\in\text{ind}(i,j)}(E_l^i-E_l^j)(E_l^i-E_l^j)^T.
\end{eqnarray}
It is straightforward to show that \eqref{Ewexpanded} is doubly stochastic with the maximum eigenvalue of 1. Note that the multiplicity of $\lambda=1$ may be more than one.
We find $\gamma$ by solving the characteristic equation associated with $\mathbb{E}[Q(k)^TQ(k)]$. From the definition of $Q(k)$ (Lemma~\ref{gamma_less_1}) and Lemma~\ref{lemma_ext_stoch} we obtain,
\begin{equation}\label{EQQ}
\mathbb{E}[Q(k)^TQ(k)]=\mathbb{E}[W(k)^TW(k)]-H\bar{H}=\bar{W}-H\bar{H}.\nonumber
\end{equation}
We solve the following characteristic equation for $\lambda$ to find $\gamma$ which is equal to $\lambda_{\max}\big(\mathbb{E}[Q(k)^TQ(k)]\big)$.
\begin{equation}\label{charac_eq_gamma}
\text{det}\Big((1-\lambda)I_m-H\bar{H}+\bar{W}-I_m\Big)=0.
\end{equation}
By the generalized Sylvester's determinant theorem, for any  $Y$ and any invertible $X$,
\begin{equation}\label{det_sum}
\text{det}(X+Y)=\text{det}(X)\text{det}(I+X^{-1}Y).
\end{equation}
Let $X:=(1-\lambda)I_m-H\bar{H}$ and $Y:=\bar{W}-I_m$. Therefore, using \eqref{charac_eq_gamma} and \eqref{det_sum}, we arrive at the following two equations:
\begin{eqnarray}
\label{det(X)}&&\text{det}((1-\lambda)I_m-H\bar{H})=0,\\
\label{det(I+X-1A)}&&\text{det}\Big(I_m-((1-\lambda)I_m-H\bar{H})^{-1}(I_m-\bar{W})\Big)=0.
\end{eqnarray}
We will solve \eqref{det(X)}, \eqref{det(I+X-1A)} after  we derive conditions for $X$ to be invertible and find $X^{-1}$.
Let $A$, $U$, $B$ and $V$ be $m\times m$, $m\times N$, $N\times N$ and $N\times m$ matrices, respectively. Let also $B$ be invertible. By the \emph{Woodbury matrix identity} we have,
\begin{eqnarray}\label{Woodbury}
\hspace{-0.2cm}&&(A\!+\!UBV)^{-1}\!=\!A^{-1}\!-\!A^{-1}U(B^{-1}\!+\!V\!A^{-1}U)^{-1}V\!A^{-1},
\end{eqnarray}
where $A$ and $B^{-1}+VA^{-1}U$ are nonsingular matrices.
Let $A:=(1-\lambda)I_m$, $U:=H$, $B:=\text{diag}(-1./\textbf{m})$ and $V:=H^T$. Thus, $X=A+UBV$. Using \eqref{Woodbury}, we find $X^{-1}$ and conditions under which $X^{-1}$ exists.
\begin{eqnarray}
&&X^{-1}=\Big((1-\lambda)I_m-H\bar{H}\Big)^{-1}\nonumber\\
&&=\Big((1-\lambda)I_m+H\text{diag}(-1./\textbf{m})H^T\Big)^{-1}\nonumber\\
&&=\frac{1}{1-\lambda}I_m-\frac{1}{(1-\lambda)^2}H\Big(\text{diag}(-\textbf{m})+\frac{H^TH}{1-\lambda}\Big)^{-1}H^T.\nonumber
\end{eqnarray}
It is straightforward to verify that the nonsingularity conditions for $A$ and $B^{-1}+VA^{-1}U$ are $\lambda\neq 1$ and $\lambda\neq 0$. Note that $H^TH=\text{diag}(\textbf{m})$ \eqref{H^TH=diag(m)}. For $\lambda\neq 0,1$ we obtain,
\begin{eqnarray}
&&X^{-1}=\frac{1}{1-\lambda}I_m-\frac{1}{(1-\lambda)^2}H\Big(\text{diag}(\textbf{m})\big(\frac{\lambda}{1-\lambda}\big)\Big)^{-1}H^T\nonumber\\
&&=\frac{1}{1-\lambda}I_m-\frac{1}{\lambda(1-\lambda)}H\bar{H}=\frac{1}{\lambda(1-\lambda)}\Big(\lambda I_m-H\bar{H}\Big).\nonumber
\end{eqnarray}
Recall \eqref{det(X)}, \eqref{det(I+X-1A)} for $\lambda$, $\lambda\neq 0,1$. Note that by Lemma~\ref{R=1}, all the solutions of \eqref{det(X)} are either 0 or 1. Thus, there is no $\lambda$ satisfying \eqref{det(X)}. Let's simplify \eqref{det(I+X-1A)} and solve it for $\lambda\neq 0,1$.
\begin{eqnarray}\label{det_tot}
&&\text{det}\Big(I_m-((1-\lambda)I_m-H\bar{H})^{-1}(I_m-\bar{W})\Big)=0\Leftrightarrow\\
&&\text{det}\Big(I_m-\frac{1}{1-\lambda}(I_m-\bar{W})+\frac{1}{\lambda(1-\lambda)}H\bar{H}(I_m-\bar{W})\Big)=0.\nonumber
\end{eqnarray}
In the following we show that $\frac{1}{\lambda(1-\lambda)}H\bar{H}(I_m-\bar{W})=\textbf{0}_N$. Using \eqref{Ewexpanded} and \eqref{Hbar}, for every $i\in V$, $j\in N_C(i)$ and $l\in\text{ind}(i,j)$ we have,
\begin{eqnarray}
\hspace{-0.2cm}&&\frac{1}{\lambda(1-\lambda)}H\bar{H}(I_m-\bar{W})=\frac{1}{2N\lambda(1-\lambda)}H\bar{H}\sum_{i\in V}\frac{1}{\text{deg}_{G_C}(i)}\nonumber\\
\hspace{-0.2cm}&&.\sum_{j\in N_C(i)}\sum_{l\in\text{ind}(i,j)}(E_l^i-E_l^j)(E_l^i-E_l^j)^T\nonumber\\
\hspace{-0.2cm}&&=\frac{1}{2N\lambda(1-\lambda)}H\text{diag}(1./\textbf{m})\sum_{i\in V}\frac{1}{\text{deg}_{G_C}(i)}\nonumber\\
\hspace{-0.2cm}&&.\sum_{j\in N_C(i)}\sum_{l\in\text{ind}(i,j)}\bigg[\sum_{i=1}^{N}{E_\tau^i}^T\bigg]_{\tau\in V}^T(E_l^i-E_l^j)(E_l^i-E_l^j)^T=\textbf{0}_N.\nonumber
\end{eqnarray}
The last equality holds because $\sum_{i=1}^{N}{E_l^i}^T(E_l^i)=\sum_{i=1}^{N}{E_l^i}^T(E_l^j)=1$ for every $i\in V$, $j\in N_C(i)$ and $l\in\text{ind}(i,j)$.
Then for $\lambda\neq 0,1$, using the definition of $\bar{W}$, \eqref{det_tot} becomes,
\begin{equation}\label{det_van}
\text{det}\Big(\!I_m\!-\!\frac{1}{1-\lambda}(\!I_m-\!\bar{W})\Big)\!=0\!\Leftrightarrow\!\text{det}\Big(\lambda I_m-\bar{W}\Big)\!=0.
\end{equation}
Note that $\gamma$ is contained in the solution set of \eqref{det_van} because $\gamma<1$ by Lemma~\ref{gamma_less_1}. Thus, we can conclude that $\gamma=\max_{\lambda\neq 1}\lambda(\bar{W}):=\lambda_2(\bar{W})$.
$\hfill\blacksquare$
\bibliographystyle{IEEEtran}
\bibliography{IEEEabrv,ref}
\end{document}